# Nanobatteries in redox-based resistive switches require extension of memristor theory

I. Valov[1,2*], E. Linn[1*], S. Tappertzhofen[1*], S. Schmelzer[1], J. van den Hurk[1], F. Lentz[2], R. Waser[1,2]

[1] Institut für Werkstoffe der Elektrotechnik II, RWTH Aachen University, 52074 Aachen, Germany

[2] Peter Grünberg Institute 7, Research Centre Jülich GmbH, 52425 Jülich, Germany

* Authors contributed equally

(Former version: Submitted on July 2nd 2012)

## Abstract

Redox-based nanoionic resistive memory cells (ReRAMs) are one of the most promising emerging nano-devices for future information technology with applications for memory, logic and neuromorphic computing. Recently, the serendipitous discovery of the link between ReRAMs and memristors and memristive devices has further intensified the research in this field. Here we show on both a theoretical and an experimental level that nanoionic-type memristive elements are inherently controlled by non-equilibrium states resulting in a nanobattery. As a result the memristor theory must be extended to fit the observed non zero-crossing *I-V* characteristics. The initial electromotive force of the nanobattery depends on the chemistry and the transport properties of the materials system but can also be introduced during ReRAM cell operations. The emf has a strong impact on the dynamic behaviour of nanoscale memories, and thus, its control is one of the key factors for future device development and accurate modelling.

## Introduction

Resistive switching memories are nanoionic-based electrochemical systems with a simple metal - ion conductor (insulator) - metal (MIM) structure. These devices exhibit low power consumption, response times in the nanosecond range, and scalability down to the atomic level[1-3]. They demonstrate excellent prospects for the application in modern information technology, in particular for novel logical devices[4] and artificial neuromorphic systems[5]. The link[6] between redox-based nanoionic resistive memories[7], memristors[8] and memristive devices[9] has further intensified the research in this important area and inspired the introduction of new concepts[10,11-13] and material systems[3, 2]. Efforts are now focused on a microscopic understanding of the physicochemical processes responsible for resistive switching[14] and on meeting the challenges of circuit design[15].

The system properties and the material performance of ReRAM devices are modulated by quantum effects, excess surface free energy of atomic clusters, and nonlinear mesoscopic transport phenomena, due to the nano-dimensions of the devices in both lateral and vertical direction, that are often comparable to space charge layer lengths. The borders of well known definitions, such as those for ion conductors and insulators, become blurred at the nanoscale - and various classes of materials starting from $RbAg_4I_5$, AgI (as bulk ion conductors) and continuing to $SiO_2$ and $Ta_2O_5$ (as bulk insulators) are all used and termed ionic or mixed ionic-electronic electrolytes at room temperature[16-20].

The cation-migration based electrochemical metallization memory cells (ECM) are a class of ReRAMs that use Ag or Cu as an active electrode and for example Pt, Ir or W as an inert counter electrode. A variety of oxide, chalcogenide, and halide thin films have been suggested for the solid electrolyte[2]. Applying a positive voltage between the active and the counter electrode leads to an oxidation (dissolution) of the active electrode's material and to a deposition of metal (Ag or Cu) at the counter electrode. Due to the high electric field in the order of $10^8$ V $m^{-1}$, the metallic deposit propagates in a filamentary form and short circuits the cell, thus defining a low-resistive ON state. The filament can be dissolved by applying a voltage of opposite polarity to return the cell to a high resistive OFF state. The anion-migration based valence change cells (VCM) typically use a high work function electrode (e. g. Pt, TiN), an oxygen-affine, lower work function electrode and a metal oxide as the electrolyte. These cells rely on the formation of oxygen deficient, mixed ionic-electronic conducting filaments and the nanoionic modification of the potential barrier between the tip of the filament and the electrode to define the ON and OFF states. The ON and OFF states are then used to read the Boolean 1 and 0, respectively.

The functionality of ReRAM cells and the kinetics of the filament formation/dissolution are the subject of intensive studies from both, academia and industry, leading to a development of empirical or semi-empirical models for the operating principles[21], expanding to include concepts of multibit memories[22] and memristive systems[10, 15, 6].

From the circuit theory's point of view, ReRAM cells are regarded as memristive elements or real memristors[23]. They are defined by two simple equations, the state-dependent Ohm's law,

$$I = G(\mathbf{x}, V) \cdot V \quad , \tag{1}$$

and the state equation

$$\frac{\mathrm{d}\mathbf{x}}{\mathrm{d}t} = \mathrm{f}(\mathbf{x}, V) \, . \tag{2}$$

The distinctive feature of memristive elements is a pinched characteristics at the origin of the *I-V* plane (i.e. the *I-V* zero-crossing property)[8, 9], which is a direct result of these equations, and therefore represents the essential fingerprint[23]. A short excursus on the use of the term 'memristor' is given in Supplementary Section S1.

Here, we report on non-equilibrium ON and OFF states in ReRAM cells determined by chemical potential gradients generating an electromotive force of up to a few hundred millivolts, violating the zero-crossing property. The emf may affect the retention time and both influences and is influenced by the processes during formation and rupture of the metallic filament. We introduce additional equations to account for the emf, i.e., for the non zero-crossing hysteresis loop, thus qualitatively and quantitatively extending the memristor theory. The conclusions we draw with respect to a series of ECM systems apply to VCM-type ReRAM cells as well as other electrochemical and (neuro-) biological systems.

## Results

To clearly identify the individual influences of different chemical potential gradients we studied ECM cells build from a series of materials, selected so as to ensure a transition of particular chemical and transport properties, i.e. $SiO_2$ – $GeS_x$, $GeSe_x$ – AgI (with x = 2.2 and 2.3, respectively). In the as-deposited state, all materials are electronic insulators where the first and the last compounds of this series represent the two extremes. That is to say, $SiO_2$ has a very small (but at the nanoscale not completely negligible) electronic conductivity and AgI is a stoichiometric ionic compound with considerable $Ag^+$ ion conductivity. Neither is able to dissolve Ag chemically. In contrast, $GeS_x$ and $GeSe_x$ are able to dissolve Ag (or Cu) to different extents moving from insulators to mixed ionic and electronic electrolytes, thus displaying a transport property transition between $SiO_2$ and AgI.

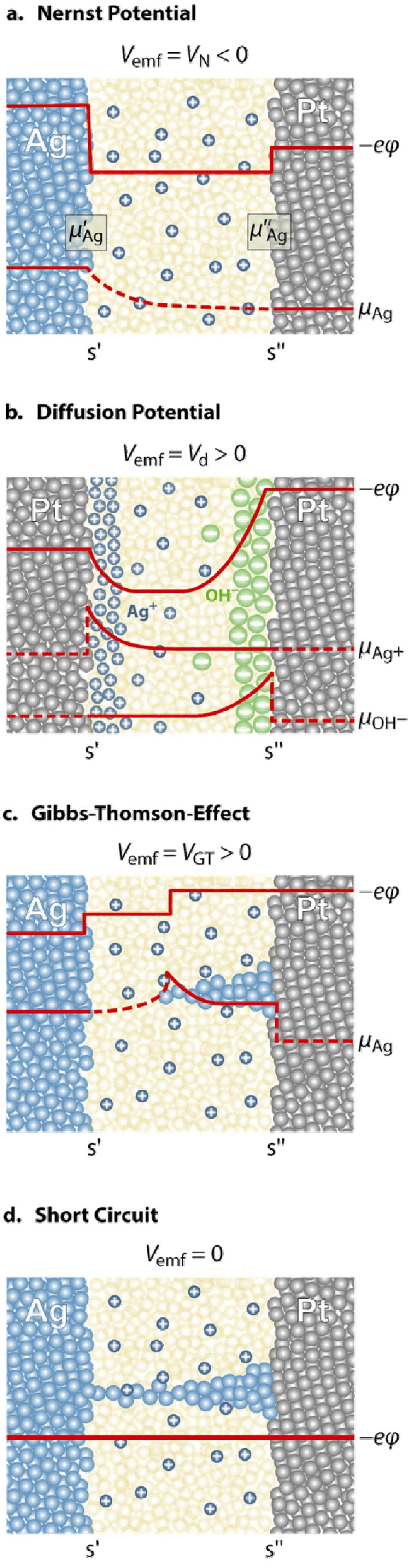

**Figure 1**. **Origins of emf in nanoscale cells.** These sketches show situations for $SiO_2$ based cells in which one of the three basis origins of emf dominates. **(a)** A Nernst potential $V_N$ of a Ag/$SiO_2$/Pt cell arising from the difference of the chemical potential of Ag metal at the interfaces Ag/electrolyte and Pt/electrolyte $\Delta\mu_{Ag} = \mu'_{Ag} - \mu''_{Ag}$. The Nernst potential $V_N$ according to equation (11) has a negative value and it is given by the difference of the electrical potentials at both electrodes $\Delta\varphi = V_{emf} = -\frac{\Delta\mu_{Ag}}{ze}$ generated to keep the condition $\sum_i \tilde{\mu}_i = 0$ ($\tilde{\mu}_i$ is the electrochemical potential given by $\tilde{\mu}_i = \mu_i + ze\varphi$). Depending on the specific chemical redox system and the chemical potential gradients, the sign of the $V_{emf}$ can also be positive. **(b)** A diffusion potential $V_d$ is generated in a Pt/$SiO_2$/Pt cell by gradients of the chemical potentials of the $Ag^+$ and $OH^-$ ions, i.e., $\Delta\mu_{Ag^+} = \mu'_{Ag^+} - \mu''_{Ag^+}$ and $\Delta\mu_{OH^-} = \mu'_{OH^-} - \mu''_{OH^-}$ inhomogeneously distributed in the thin film as given by equation (5). **(c)** In the case of a nanosize filament, the chemical potential of Ag contains an additional surface energy term generating a chemical potential gradient $\Delta\mu_{Ag} = \mu_{Ag\text{-micro}} - \mu_{Ag\text{-nano}}$ in accordance with equation (6) (Gibbs-Thompson Potential $V_{GT}$). **(d)** In the case of a fully metallic contact or a highly conducting tunnel junction, the emf is $V_{cell} = 0$ . The potential of the right electrode is used as a reference throughout the text. Please note that profiles of the electrostatic potential $\varphi$ is sketched without zooming into the space charge layers.

We found three factors which contribute to the formation of the cell voltage $V_{\mathrm{emf}}$, as illustrated in Fig. 1: 1. the classical Nernst potential $V_{\mathrm{N}}$ (Fig. 1a), 2. the diffusion potential $V_{\mathrm{d}}$ (Fig. 1b), and 3. the Gibbs-Thompson potential due to the different surface free energies of macro- and nanoparticles $V_{\mathrm{GT}}$ (Fig. 1c), for the case that a metallic nanofilament is formed without short-circuiting the electrodes. Note that in the case that a metallic nanofilament is formed, the emf is zero due to the short circuit (Fig. 1d).

The Nernst voltage $V_{\mathrm{N}}$ is given by the difference between the potential-determining half-cell reactions at each electrode/electrolyte interface:

$$V_{\mathrm{N}} = V_{\mathrm{s'}} - V_{\mathrm{s''}} = V^0 + \frac{kT}{ze} \ln \frac{(a_{\mathrm{Me}^{z+}})_{\mathrm{s'}} \cdot (a_{\mathrm{Red}})_{\mathrm{s''}}}{(a_{\mathrm{Me}})_{\mathrm{s'}} \cdot (a_{\mathrm{Ox}})_{\mathrm{s''}}} \tag{3}$$

with $V_{\mathrm{s'}}$ and $V_{\mathrm{s''}}$ being the half-cell potentials at the active electrode/electrolyte (s') and inert electrode/electrolyte (s'') interfaces, $z$ the number of exchanged electrons, and $V^0$ the difference in the standard potentials of these reactions. At the s' interface, the potential-determining reaction is the same for all ECM cells:

$$\mathrm{Me}^{z+} + ze^{-} \rightleftharpoons \mathrm{Me} \tag{4}$$

At the s'' interface, Red and Ox are general expressions in eq. (3) for a species undergoing a redox process. The nature of this species is determined by the specific cell, as will be shown below.

The non-equilibrium diffusion potential $V_{\mathrm{d}}$ in ECM cells arises due to excess concentrations of charged species i.e., $Ag^{+}$ ions, electrons, and/or $OH^{-}$ ions within the electrolyte film. These ions are introduced into the solid films either electrochemically during the forming of SET/RESET cycles or chemically due to a chemical dissolution of Ag. In both cases, the ratio of concentration changes across the electrolyte layer, with particularly pronounced changes in the vicinity of the electrodes. The electromotive force generated by this inhomogeneous charge distribution and mobilities is given by[24]:

$$V_{\mathrm{d}} = -\frac{kT}{e} \sum_i \int_{\mathrm{s''}}^{\mathrm{s'}} \frac{t_i}{z_i} \, d \ln a_i = -\frac{kT}{e} \left( \overline{t}_{\mathrm{Me}^+} \ln \frac{(a_{\mathrm{Me}^+})_{\mathrm{s'}}}{(a_{\mathrm{Me}^+})_{\mathrm{s''}}} - \overline{t}_{-} \ln \frac{(a_{-})_{\mathrm{s'}}}{(a_{-})_{\mathrm{s''}}} \right) \tag{5}$$

where $k$, $T$ and $e$ are the Boltzmann constant, the temperature, and the elementary charge, respectively; $t_i$ denotes the transference number of the species $i$ (with metal cation $Me^{+}$ and "$-$" representing anions and electrons), $z_i$ their charge number, and $a_i$ their activities at the electrolyte sides of the interfaces s' and s'' which correspond to the active electrode/electrolyte and the inert electrode/electrolyte interfaces, respectively. The potential difference generated in the neuron cells to transport electric signals has the same nature and originates in the diffusion (Donnan) potential[24].

The contribution of the third component $V_{GT}$ to the total cell voltage $V_{emf}$ is expected in the case of a noncontacting filament due to the different surface free energies of the macrocrystalline active electrode and the nanosize filament in accordance with the Gibbs-Thomson equation:

$$V_{GT} = -\frac{\mu_{Ag}^{macro} - \mu_{Ag}^{nano}}{ze} = -\frac{2\gamma}{zer}V_m, \qquad (6)$$

where $\gamma$ is the surface free energy, $r$ is the radius of the particle, $V_m$ is the molar volume and $\mu_{Ag}$ is the chemical potential of Ag. $V_{GT}$ can be observed, for instance, for highly ohmic ($R$ > 12.9 kΩ) ON states (no metallic short circuit) typically caused by filaments in ECM cells without galvanic contact to the active electrode (Supplementary Section S2). It should be noted that the contribution of $V_{GT}$ could not be clearly distinguished from the contribution of $V_d$ in the particular systems we have studied.

In Fig. 2 we show the time evolution of the open cell voltage $V_{cell}$ and of the short circuit current $I$ for $Ag/SiO_2/Pt$ cells. Fig. 2a displays a simplified equivalent circuit of a battery in which the cell voltage

$$V_{cell} = t_{ion}V_{emf} = \frac{R_i^{-1}}{R_i^{-1} + R_e^{-1}}V_{emf} \qquad (7)$$

where $t_{ion}$ represents the total ionic transfer number, $R_i$ is the total resistance of the ionic current path and $R_e$ the total electronic resistance of the cell.

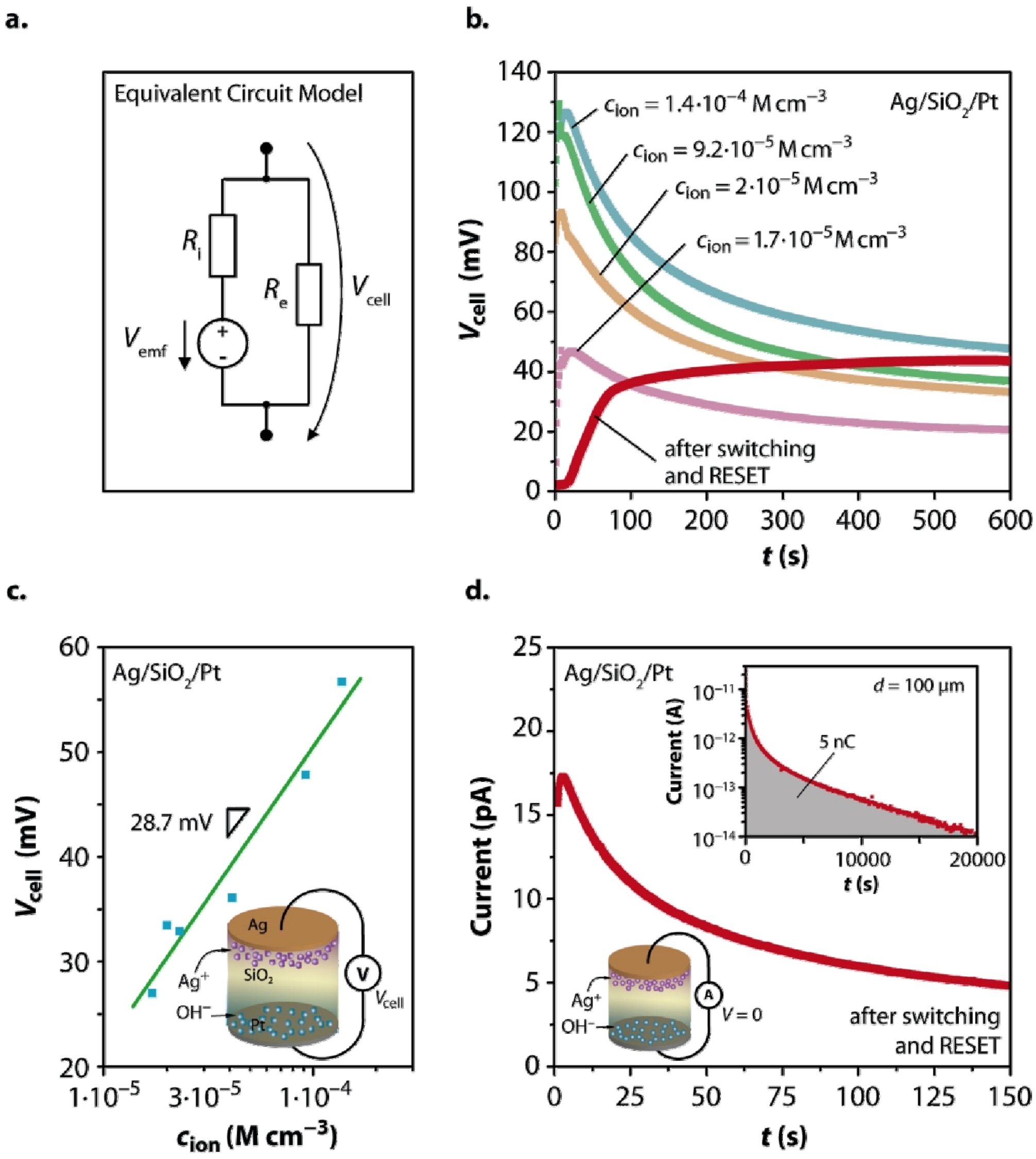

**Figure 2**. **Steady state emf measurements for Ag/SiO$_2$/Pt ECM cells.**

**(a)** Simplified equivalent circuit model of a ReRAM device. **(b)** $V_{cell}$ for a Ag/SiO$_2$/Pt cell measured under open circuit conditions. The red line depicts the $V_{cell}$ in the OFF state after a SET/RESET cycle. For the other curves, the ion concentration $c_{ion}$ (i.e. the sum of the $Ag^+$ and $OH^-$ ion concentrations, averaged over the thickness) was controlled and preset using different sweep rates[25]. Details of the measurements can be found in Supplementary Section S3. **(c)** The slope of the line provides the pre-exponential term and we were thus able to determine the ionic transference number $\overline{t}_{ion}$ by using eq. (10) and $z = +1$ (for $Ag^+$). **(d)** The time evolution of the discharge current (for $V = 0$ V) of the cell is shown. Inset: the same plot for an extended time and a log current scale. The integration reveals the charge of the nanobattery.

Fig. 2b shows $V_{cell}$ after establishing a defined concentration $c_{ion}$ of $Ag^+$ ions for pristine cells. We utilized the fact that the amount of ions generated at the s' and s'' interfaces during cell operation is adjusted by the pulse length and height or the sweep rate[25]. The decrease of $V_{cell}$ over a few hundred seconds is caused by the equilibration of the $Ag^+$ concentration gradient which has an initial maximum immediately at the metal/electrolyte interface. Furthermore, Fig. 2b (red line) shows $V_{cell}$ immediately after a RESET operation and switching to the open cell measuring conditions. In this case the $Ag^+$ concentration in the immediate vicinity to the Ag electrode is depleted (in contrast to pristine cells) on the one hand side due to the reduction process (i.e. the RESET process) and on the other hand side due to slow supply/diffusion of ions from layers apart from the interface (diffusion limited). Thus, in this situation $V_{cell}$ increases with time but relaxes to the same value as the formed pristine cells after equilibration.

The potential-determining reaction at the inert electrode e.g. Pt depends on the material properties of the solid electrolyte and we distinguish two cases. The system Ag/$SiO_2$/Pt represents the first case for which the solid $SiO_2$ film contains no initial $Ag^+$ and, hence, reaction (4) cannot be potential-determining. Instead, as shown in Ref. [26], moisture is typically incorporated into this electrolyte during fabrication and hence protons provide the required counter reaction at the inert electrode e.g.,

$$\tfrac{1}{2}O_2 + H_2O + 2e^- \rightleftharpoons 2OH^- \ \left(\text{or alternatively } 2H_2O + 2e^- \rightleftharpoons 2OH^- + H_2\right) \tag{8}$$

and the Nernst voltage takes the form:

$$V_N = V^0 + \frac{kT}{2e} \ln \frac{(a^2_{Me^{z+}})_{s'} \cdot (a^2_{OH^-})_{s''}}{(a^2_{Me})_{s'} \cdot (a^{1/2}_{O_2})_{s''} \cdot (a_{H_2O})_{s''}} \tag{9}$$

The total emf of the ECM cell is a combination of equations (5) and (9) and is expressed by:

$$V_{emf} = V_N + V_d = V_0 + \bar{t}_{OH^-} \frac{kT}{e} \ln(a_{Me^+})_{s'} + \bar{t}_{Me^+} \frac{kT}{e} \ln(a_{OH^-})_{s''} \sim V_0 + \bar{t}_{ion} \frac{kT}{2e} \ln(a_{ion}) \tag{10}$$

with $V_0 = V^0$ + const. (see Supplementary Section S4).

In Fig. 2c we evaluate $V_{cell}$ after initial relaxation. The slope of the line provides the pre-exponential term and we were thus able to determine the ionic transference number $\bar{t}_{ion} = \bar{t}_{Ag^+} + \bar{t}_{OH^-} = 0.4$ by using eq. (10). The x-axes intercept corresponds to the condition $\ln(a_{ion}) = 0$ and provides the value of $V_0 = 0.17$ V. Both values are used as parameters in the device modeling below.

Fig. 2d illustrates the short-circuit currents of a Ag/$SiO_2$/Pt cell after RESET operation. In the inset, an extended discharge time of 20000 s (approx. 5.5 h) is shown. The measurements were carried out until they approached the resolution limit of our measuring system (approx. 10 fA). The current is proportional to the electrode area (see Supplementary Section S5). An integration of current over time reveals a total charge of approx. 5 nC released during the discharge of the battery which corresponds to a conversion of approximately 2 monolayers Ag. In comparison, the dielectric discharge of the cell capacitor is approx. 0.4 pC, clearly demonstrating that the discharge phenomenon is of an electrochemical nature.

Please note that retention time of the cell (here: in the OFF state) is not immediately related to the relaxation time of the emf voltage. While the first has to be > 10 years for universal non-volatile memories, the latter may be in the order of minutes to days as shown. In other words, the discharge of the ReRAM nanobattery does not lead to a loss of the OFF state (in similar manner as the discharge of any other rechargeable battery does to end up in a filamentary short between the electrodes). Instead, typically only a shift of the particular $R_{OFF}$ value is observed upon the relaxation of the emf. Such a shift has been reported e.g. by [27] and [28] for both ECM and VCM systems, respectively. More detailed discussion is provided in Supplementary Section S6.

In the second case condition for establishing a Nernst voltage according to eq. (3), the electrolyte contains mobile $Me^{z+}$ ions dissolved by electrochemical and/or chemical processes with almost homogeneous distribution i.e., $(a_{Me^{z+}})_{s'} \sim (a_{Me^{z+}})_{s''}$. Then the potential-determining half-cell reaction at the s'' interface will be the same as at s', i.e. reaction (4), and, because no chemical potential gradient of the charged species is assumed, equation (3) can be simplified to:

$$V_{emf} = \bar{t}_{ion} \frac{kT}{e} \ln \frac{(a_{Me})_{s''}}{(a_{Me})_{s'}} \tag{11}$$

where $(a_{Me})_{s'}$ denotes the activity of the active metal Me at interface s', i.e. $(a_{Me})_{s'} = 1$, if a pure metal is used, and $(a_{Me})_{s''}$ is the activity of Me a the inert counter electrode s'', i.e. $(a_{Me})_{s''}$ is typically very low. $\bar{t}_{ion}$ is the ion transference number averaged throughout the electrolyte thickness. Because $(a_{Me})_{s'}$ is fixed, the Nernst voltage is a function of $(a_{Me})_{s''}$ alone. We succeeded in demonstrating a Nernst emf in accordance with equation (11) (Fig. 1a) for the systems Ag/Ag-$GeS_{2.2}$/Pt and Ag/Ag-$GeSe_{2.3}$/Pt which contain a significant amount of chemically dissolved $Ag^+$ ions (a solubility of up to 35 at% Ag has been reported[29]). We applied a positive external voltage of 1 V to the inert electrode, thus removing the residual Ag atoms from the s'' interface and lowering their activity. In the coupled cathodic reaction at the Ag interface s', Ag dendrites are deposited as shown in the optical image in Fig. 3.

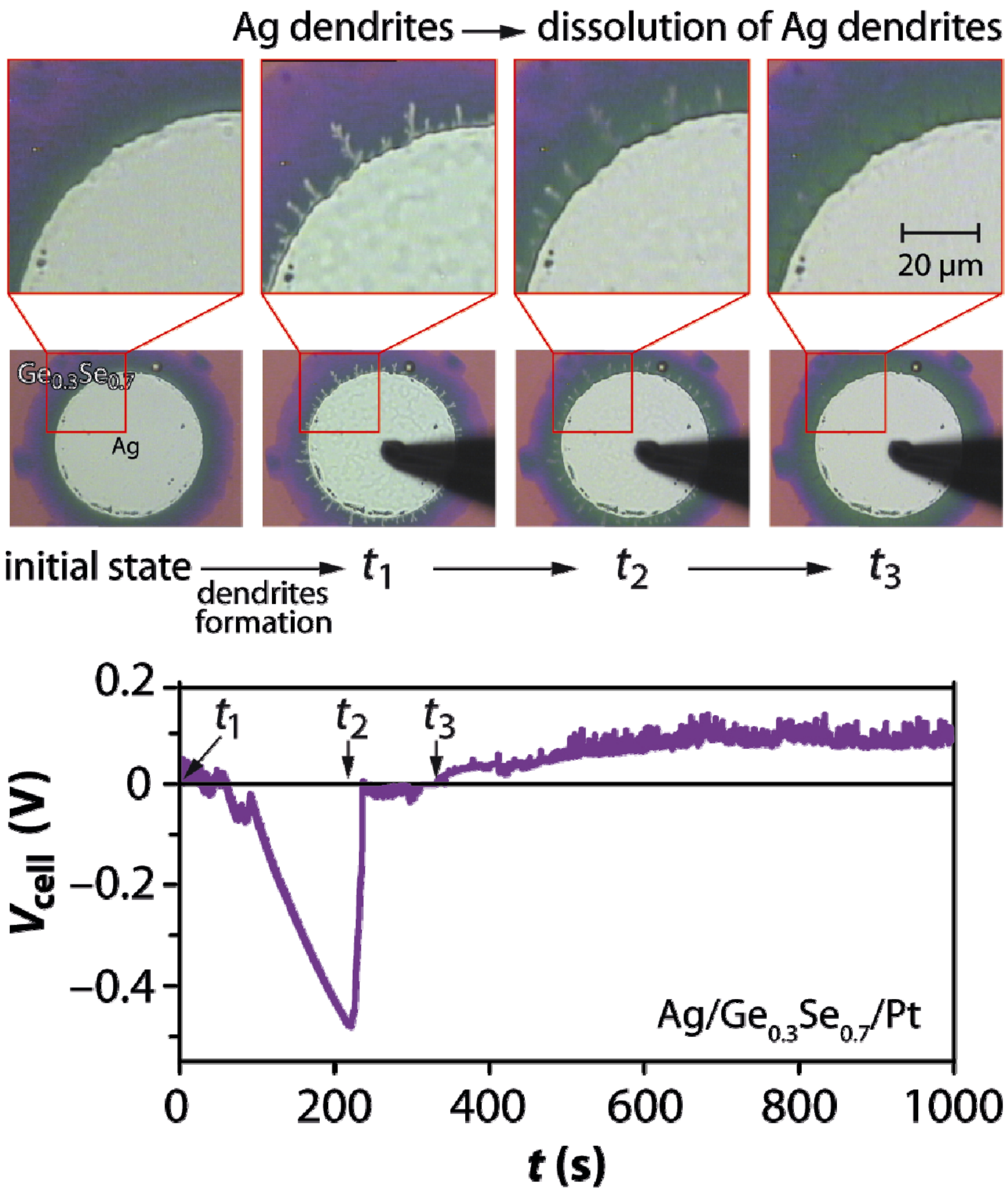

**Figure 3. Time dependent emf measurements for Ag/$Ge_{0.3}Se_{0.7}$/Pt ECM cells.**

After a SET/RESET cycle (initial state) a negative voltage is applied to the Ag electrode resulting in a dendrite formation. At $t_1$ = 0s the voltage measurement is started. Fading of the diffusion potential contribution from $t_1$ to $t_2$ leads to a highly negative $V_{cell}$ due to a significant Nernst potential between the electrodes. In parallel to the dissolution of the dendrites ($t_2$) the emf increases to positive values again with a diffusion potential as the remaining component of the emf due to the different activity of $Ag^+$ ions at the both s', s'' surfaces ($t > t_3$).

Thus, apart from the different concentration of $Ag^+$ ions at both interfaces we generated a difference in the activities of Ag. The emf of the system was then monitored over time and correlated to the evolution of the dendrite morphology further recorded by optical images. The time evolution shown in Fig. 3 arises from an interplay of a Nernst potential and a diffusion potential. In the voltage modified state ($t_1$ in Fig. 3), the contribution of the (positive) diffusion potential fades, leading to a dominating (negative) Nernst potential, leading to an emf value of approx. –450 mV due to the contribution of eq. (11). The emf value corresponds to a difference in the Ag activities at both electrodes, $(a_{Ag})_{s''} \approx 2.5 \cdot 10^{-8} (a_{Ag})_{s'}$. As monolayer(s) of Ag begin to form at the Pt electrode (in parallel to the dissolution of the dendrites at the Ag electrode) the emf increases again simultaneously and relatively fast because of $(a_{Ag})_{s'} \approx (a_{Ag})_{s''}$. But due to slower diffusion of the $Ag^+$ ions $a(Ag^+)_{s'} > a(Ag^+)_{s''}$ the positive diffusion potential governs $V_{cell}$. Thus, the control over $V_{cell}$ changes from a situation mainly determined by equation (11) to a situation dominated by equation (5).

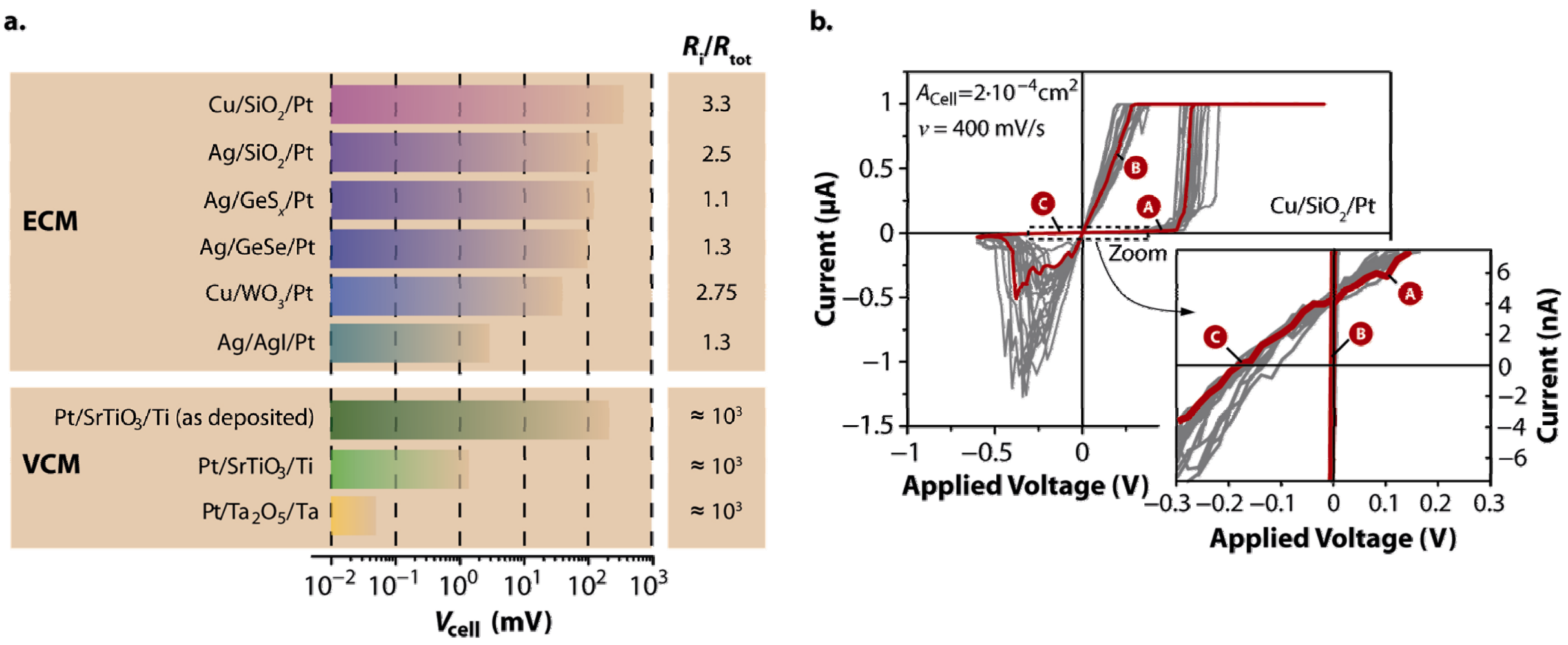

**Figure 4**. **Emf for different types of ReRAM cells.** All tested cells showed an emf varying from some hundreds of microvolts to some hundreds of millivolts **(a).** The specific steady state emf depends on the type of cell and the device operation (e.g. sweep rate or voltage amplitude prior to the (open cell) emf measurement). For the sake of completeness, the internal resistance $R_i$ normalized by the total resistance $R_{tot}$ is given (for details: see Supplementary Section S11). The influence of the emf on the current-voltage sweeps, resulting in non zero-crossing characteristics **(b),** is given for the $Cu/SiO_2/Pt$ system (red curve) as an example. The *I-V* characteristics in grey are provided for statistical verification. For the sake of clarity, the currents of the OFF state (A) and (C) and the ON state (B) are labelled, respectively. The formed valence change memory (VCM)-type cells ($Ti/SrTiO_3/Pt$ and $Ta/Ta_2O_5/Pt$) show lower emf for the OFF state due to the higher electronic partial conductivity i.e. higher currents, but lower ion transference number. Details of VCM and also of $Ag/SiO_2/Pt$ cells are presented in Supplementary Section S8 and S9, respectively.

In the course of our emf studies, we investigated the series of different ECM cells as well as selected VCM cells. We were able to prove that in all types of ReRAM cells tested significant (electro)chemical potential gradients are generated by the operation of the cells. Inevitably, these gradients give rise to an emf, and, hence, the cells show the characteristics of nanosize batteries with cell voltages in the range displayed in Fig. 4a (for details see Supplementary Section S11). After forming, the cell voltages of VCM cells are much lower than for ECM cells due to the significantly higher electronic conductance in the OFF state originating from the stub of the highly conductive filament.

**Discussion**

The non-equilibrium states reflected by the emf voltage may affect both the retention and the device operation. Whereas OFF and intermediate states always experience emf voltages, for ON states with metallic contact the cell voltage is generally close to zero due to the electronic conductance of the filament. Apart from that, chemical potential gradients, size effects and electrolyte non-stoichiometry can result in a chemical dissolution of the filament [30]. Thus, the retention of the ON state is strongly dependent on the filament characteristics and only indirectly dependent on the emf voltage.

Being an intensive state property, the $V_{emf}$ is independent on the cell size. However, the ionic and electronic resistances, $R_i$ and $R_e$, and hence $V_{cell}$ may scale in a non-trivial manner depending on the ReRAM type, e.g. the properties of the filament stub and the gap between the filament tip and the electrode in the case of VCM cells. A more detailed discussion of this point is presented in Supplementary Section S7.

Our results strongly suggest that in all bipolar ReRAM devices a nanobattery with dedicated emf voltages is present. This fact has far-reaching consequences on the application of the theory of memristive elements and on the device modelling in general. The emf is a state property corresponding to a non zero-crossing $I$-$V$-characteristic (compare Fig. 4b), and we correspondingly expanded the memristive equations to obtain an active device which can be considered an extended memristive element. In fact, any electrochemical system is active by nature, thus two-terminal nanoionic resistive switches cannot be pure passive memristors. The same is true for neurobiological systems, which are also active, offering passive memristive elements only as internal elements[31]. However, in contrast to the emf of ReRAM cells, the biological resting membrane potential is modulated only by $V_d$ (eq. (5)) and not by $V_N$ (eq. (3)). By abandoning the zero-crossing property, a large number of dynamical devices can be added to the framework of memristive, memcapacitive and meminductive devices. For example, ferroelectric capacitors as well as ferromagnetic inductors can thus be regarded as extended memcapacitive and meminductive elements, respectively (Fig. 5a). By means of this expanded framework, a ReRAM cell is still a memristive device, as shown in Fig. 5a.

The starting point for a practical memristive model of a ReRAM cell is the memristive model from Ref. [6], to which we add the nanobattery in parallel. The resulting model, which we term an extended memristive model, consists of a voltage source $V_{\mathrm{emf}}$ and a nonlinear internal resistance $R_{\mathrm{i}}$ – which together form the nanobattery – and a parallel resistor $R_{\mathrm{el}}$ whose state variable $x$ is controlled by the nanobattery (Fig. 5b and Supplementary Section S10). The state-dependent resistance of the electronic current path ($R_{\mathrm{el}}$) is a non-linear function of the applied voltage, for example a tunnelling equation[32]. The electronic leakage current is accounted for by a further parallel resistance $R_{\mathrm{leak}}$ which is state-independent and already present in the pristine device. Both contributions, $R_{\mathrm{el}}$ and $R_{\mathrm{leak}}$ in parallel represent $R_{\mathrm{e}}$ in Fig. 2a. The resistance of the ionic current path ($R_{\mathrm{i}}$) is defined by another non-linear equation, determined by the Butler-Volmer equation and/or the high-field drift equation. The state-dependent Ohm's law for the extended memristive device reads:

$$I = I_{\mathrm{ion}}\left(V_{\mathrm{emf}}, V\right) + I_{\mathrm{el}}\left(x, V\right) = G\left(x, V\right)\cdot\left(V - t_{\mathrm{ion}}V_{\mathrm{emf}}\right) \tag{12}$$

and the state equation (compare Fig. 5b) reads:

$$\dot{x} = K_1 \cdot I_{\mathrm{ion}} \quad \text{with} \quad 0 \le x \le d \tag{13}$$

where $d$ is the thickness of the active layer and $K_1$ is a constant while $I_{\mathrm{ion}}$ offers a highly non-linear voltage dependence, responsible for the pronounced nonlinearity of the switching kinetics.

To fit the observed changes in the emf measurements (Fig. 2), a second state variable, the ion concentration $c_{\mathrm{ion}}$, is required. In this case, the emf equation (10) with the experimentally determined $V_0$ = 0.17 V for Ag/$SiO_2$/Pt (Fig. 2c) reads:

$$V_{\mathrm{emf}} = V_0 + \frac{kT}{2e}\ln\left(\frac{c_{\mathrm{ion}}}{c_0}\right) \tag{14}$$

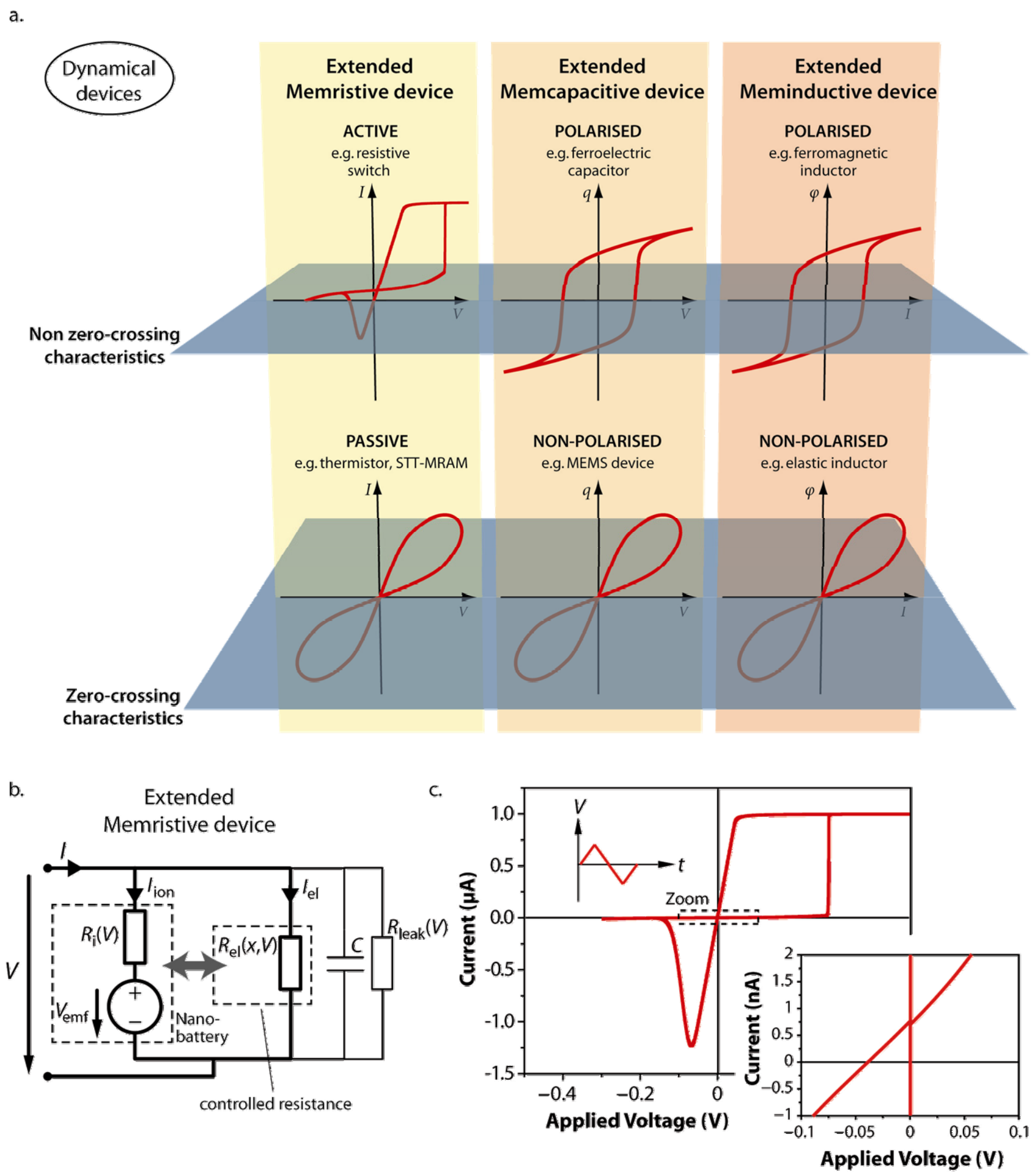

**Figure. 5 Classification of mem-devices. (a)** Memristive, memcapacitive and meminductive devices are assumed to offer pinched characteristics at the origin in general[39]. In the case of memcapacitive and meminductive devices, a spontaneous polarization of a ferroelectric and ferromagnetic material, respectively, leads for example, to non zero-crossing characteristics[11]. Similarly, a non zero-crossing *I*-*V* characteristic of a memristive device indicates the presence of an inherent nanobattery, i.e., this device is active. Therefore, we introduce the generic terms extended memristive device, extended memcapacitive device and extended meminductive device to account for both zero-crossing and non zero-crossing*I*-*V* characteristics. Note that the origin of non zero-crossing behavior in memcapacitive and meminductive devices is of completely different nature than in memristive devices, thus require specific modifications of the concept. Interestingly, Spin-Transfer Torque (STT) MRAM cells offer zero-crossing in contrast to ReRAM cells, thus can be considered as conventional memristive device. **(b)** Equivalent circuit of the extended memristive element. The ionic current is defined by the nanobattery which controls the state-dependent resistor representing the electronic current path. The capacitance of the device is neglected since its influence is not significant. The partial electronic conductivity in the electrolyte induces a state-independent resistance $R_{\text{leak}}$ due to a leakage current in parallel. (see Supplementary Section S10) **(c)** Simulated *I*-*V* characteristic of the extended memristor. The zoom shows the non zero-crossing behaviour. Further considerations on the simulation are described in Supplementary Section S10.

Modelling details are described in the Supplementary Section S10 and corresponding simulation results are depicted in Fig. 5c. The inset clearly shows the non-zero-crossing *I-V* behaviour. Thus, ReRAM cells are non zero-crossing devices, and therefore, the original memristor theory must be significantly extended in order to accommodate redox-based resistive switching systems.

Besides modeling accuracy, the emf also has direct impact on future memory device development and corresponding circuitry. Firstly, the stability of intermediate resistive states (e.g. required in multilevel memory and neuromorphic applications) must be carefully considered in terms of the emf. Moreover, the effect of the emf within ultra-dense passive crossbar arrays may become relevant due to possible device-to-device interactions. Secondly, the READ voltage of future generation memory elements will be in the range of 100 mV to 200 mV (Ref. [33]). This voltage is in the same order of magnitude as the emf voltage of some of the ReRAM types studied here. Thirdly, the READ current for a ReRAM cell in a memory matrix will be in order of 100 nA and the length of the READ pulse less than 100 ns. Hence, during the READ operation a total charge of Q = 100 nA ·100 ns = 10 fC will be transferred to or from a memory cell. As shown in our paper, the total nanobattery capacity for scaled ReRAM cells may easily be in the same order of magnitude or higher. These points demonstrate impressively the relevance of nanobattery effect for the performance and reliability of future ReRAM devices, and the necessity of including it in the device modeling. Apart from that, the emf allows for novel read-out approaches in ReRAM cells, for example, we suggested using the emf for non-destructive read-out of complementary resistive switches[34]. The theoretical implication has been raised in Refs. [35, 36].

In conclusion, we found clear indications that non-equilibrium states occur in nanoionic ReRAM cells which are generated by chemical processes such as the dissolution of the active electrode material into the electrolyte, by electrochemical processes, and by charge redistribution during the operation of the cells. The resulting electromotive force suggests the presence of a nanobattery inside the memristive device as an inherent property of the system. The nature of ReRAM systems implies a mandatory extension of the memristor theory, in order to include the non zero-crossing characteristics.

## Methods

*Sample preparation*

All samples were prepared using platinised silicon wafers as substrates. The solid electrolytes/insulators were deposited on the substrate using e-beam or thermal evaporation or RF sputtering.

**$SiO_2$**. 30 nm to 50 nm $SiO_2$ films were deposited by electron beam evaporation at $10^{-4}$ Pa at a deposition rate of 0.01 nm $s^{-1}$. Afterwards, pattern transfer of top microelectrodes with a diameter of 100 μm to 250 μm was done using conventional UV lithography. Ag (Cu) with a thickness of 30 nm was deposited by e-beam evaporation (evaporation speed 0.025 nm $s^{-1}$) followed by a 100 nm thick DC-sputtered Pt layer to prevent chemical oxidation of the silver (copper) electrode in air.

**$WO_{3-x}$.** 30 nm $WO_{3-x}$ films were sputtered in Ar plasma at a pressure of 10 Pa with an RF power of 121 W, using a $WO_3$ target. Subsequently, 70 nm thick Cu electrodes 50 μm in diameter (UV lithography) were thermally evaporated.

**$GeS_{2.2}$**. 50 nm $GeS_{2.2}$ films were deposited by RF sputtering at a gas flow of 50 sccm, 20 Pa Ar pressure and RF power of 20 W. The S/Ge ratio was determined by EDX (Oxford Instruments ISI 300, at 20 kV acceleration voltage). Ag electrodes, microstructured by UV lithography, with diameters between 10 μm and 100 μm and a thickness of 100 nm were RF sputtered at a gas flow of 50 sccm, RF power of 50 W and 20 Pa pressure. The relatively high pressure in the deposition chamber allows the deposition of homogeneous thin films[37].

**$GeSe_{2.3}$**. 50 nm to 70 nm $GeSe_{2.3}$ films were sputtered at a gas flow of 40 sccm, process pressure of $7.10^{-2}$ Pa and RF power of 13 W. The deposition rate was 0.075 nm $s^{-1}$. The preparation of the electrodes is identical to that described for $GeS_{2.2}$.

**AgI**. 30 to 50 nm AgI films were prepared by thermal evaporation. To avoid any impact of UV light and chemicals, the pattern transfer for the top electrode was done directly after subtractive pattern transfer of the Pt bottom electrode using reactive ion etching. A detailed description of the fabrication of Ag/AgI/Pt cells can be found in Ref. 17.

**$SrTiO_3$**. 70 nm of Ti was sputtered in argon plasma at an RF power of 20 W, a pressure of 0.8 Pa and a substrate temperature of 200 °C, followed by the application of 8 nm of $SrTiO_3$, using the same process parameters and a sintered ceramic target. The Pt top electrodes with a thickness of 30 nm were sputtered at a pressure of 0.8 Pa and RF power of 12 W for lower kinetic impact. All layers were deposited without breaking the vacuum to prevent contamination.

**$Ta_2O_5$.** 50 nm of Ta was sputtered in argon plasma at an RF power of 14 W, a pressure of 0.8 Pa and a substrate temperature of 200 °C. Subsequently, the Ta layer was oxidized using an argon-oxygen atmosphere with a ratio of 3:1 and an oxygen partial pressure of 100 Pa, forming a $Ta_2O_5$ surface layer with a thickness of 15-20 nm. The Pt top electrodes with a thickness of 30 nm were sputtered at a pressure of 0.8 Pa and RF power of 12 W for lower kinetic impact. All layers were deposited without breaking the vacuum to prevent contamination.

*Electrical characterization*

The emf measurements were performed in a four-needle electrode microprobe station equipped with micromanipulators to contact the sample electrodes and an optical microscope to visually monitor the surface. For potentiodynamic current-voltage measurements (CV), we applied triangular voltage sweeps between -1.5 V and 1.5 V at various sweep rates (30 mV/s to 2 V/s) using a Keithley 6430 sub-femtoampere SourceMeter. Details of the measuring method can be found in refs 28, 29. Electromotive forces were measured using the 6430 SourceMeter with high input impedance (> $10^{14}$ Ω), a Keithley 617 electrometer (> 200 TΩ) and a Keithley 2636A SourceMeter (> $10^{14}$ Ω) for comparison. Throughout the paper, the right electrode in cells denoted M'/E/M" was used as reference electrode for all measurements. We used triaxial cables and electrostatic shielding to avoid RFI effects. The offset voltage measured across a 10 MΩ resistor has been proven to be within the device specification (below 1 μV accuracy)[38]. Details on the measurement resolution and accuracy can be found in the Supplementary Section S12.

**References**:

1. Aono, M. & Hasegawa, T. The Atomic Switch. *Proc. IEEE* **98**, 2228-2236 (2010).

2. Valov, I., Waser, R., Jameson, J. R. & Kozicki, M. N. Electrochemical metallization memories-fundamentals, applications, prospects. *Nanotechnology* **22**, 254003/1-22 (2011).

3. Waser, R., Dittmann, R., Staikov, G. & Szot, K. Redox-Based Resistive Switching Memories - Nanoionic Mechanisms, Prospects, and Challenges. *Adv. Mater.* **21**, 2632-2663 (2009).

4. Borghetti, J., Snider, G. S., Kuekes, P. J., Yang, J. J., Stewart, D. R. & Williams, R. S. 'Memristive' switches enable 'stateful' logic operations via material implication. *Nature* **464**, 873-876 (2010).

5. Fölling, S., Türel, Ö. & Likharev, K. Single-electron latching switches as nanoscale synapses. *Proc. IJCNN '01*, 216-221 (2001).

6. Strukov, D. B., Snider, G. S., Stewart, D. R. & Williams, R. S. The missing memristor found. *Nature* **453**, 80-83 (2008).

7. Waser, R. & Aono, M. Nanoionics-based resistive switching memories. *Nat. Mater.* **6**, 833-840 (2007).

8. Chua, L.O. Memristor-the missing circuit element. *IEEE Trans. Circuit Theory* **CT-18**, 507-519 (1971).

9. Chua, L.O. & Kang, S.M. Memristive devices and systems. *Proc. IEEE* **64**, 209-223 (1976).

10. Ohno, T., Hasegawa, T., Tsuruoka, T., Terabe, K., Gimzewski, J. K. & Aono, M. Short-term plasticity and long-term potentiation mimicked in single inorganic synapses. *Nat. Mater.* **10**, 591-595 (2011).

11. Pershin, Y. V. & Di Ventra, M. Memory effects in complex materials and nanoscale systems. *Adv. Phys.* **60**, 145-227 (2011).

12. Yang, J. J., Strukov, D. B. & Stewart, D. R. Memristive devices for computing. *Nat. Nanotechnol.* **8**, 13-24 (2013).

13. Mazumder, P., Kang, S. & Waser, R. Memristors: Devices, Models, and Applications - Scanning the Issue. *Proceedings of the IEEE* **100**, 1911-1919 (2012).

14. Valov, I. *et al.* Atomically controlled electrochemical nucleation at superionic solid electrolyte surfaces. *Nat. Mater.* **11**, 530-535 (2012).

15. Linn, E., Rosezin, R., Kügeler, C. & Waser, R. Complementary Resistive Switches for Passive Nanocrossbar Memories. *Nat. Mater.* **9**, 403-406 (2010).

16. Yang, B., Liang, X. F., Guo, H. X., Yin, K. B., Yin, J. & Liu, Z. G. Characterization of $RbAg_4I_5$ films prepared by pulsed laser deposition. *J. Phys. D-Appl. Phys.* **41**, 115304/1-5 (2008).

17. Ilia Valov, & Kozicki, Michael N. Cation-based resistance change memory. *J. Phys. D Appl. Phys.* **46**, 074005 (2013).

18. Tappertzhofen, S., Valov, I. & Waser, R. Quantum conductance and switching kinetics of AgI based microcrossbar cells. *Nanotechnology* **23**, 145703 (2012).

19. Yao, J. *et al.* Resistive Switching in Nanogap Systems on $SiO_2$ Substrates. *Small* **5**, 2910-2915 (2009).

20. Sakamoto, T., Lister, K., Banno, N., Hasegawa, T., Terabe, K. & Aono, M. Electronic transport in $Ta_2O_5$ resistive switch. *Appl. Phys. Lett.* **91**, 092110 (2007).

21. Jameson, J. R. *et al.* One-dimensional model of the programming kinetics of conductive-bridge memory cells. *Appl. Phys. Lett.* **99**, 063506 (2011).

22. Russo, U., Kamalanathan, D., Ielmini, D., Lacaita, A. L. & Kozicki, M. N. Study of Multilevel Programming in Programmable Metallization Cell (PMC) Memory. *IEEE Trans. Electron Devices* **56**, 1040-1047 (2009).

23. Chua, L.O. Resistance switching memories are memristors. *Appl. Phys. A-Mater. Sci. Process.* **102**, 765-783 (2011).

24. Vetter, Klaus J. *Electrochemical kinetics* (Springer Verlag, 1961).

25. Tappertzhofen, S., Mündelein, H., Valov, I. & Waser, R. Nanoionic transport and electrochemical reactions in resistively switching silicon dioxide. *Nanoscale* **4**, 3040-3043 (2012).

26. Tsuruoka, T., Terabe, K., Hasegawa, T., Valov, I., Waser, R. & Aono, M. Effects of Moisture on the Switching Characteristics of Oxide-Based, Gapless-Type Atomic Switches. *Advanced Functional Materials* **22**, 70-77 (2012).

27. Miao, F., Yang, J. J., Borghetti, J., Medeiros-Ribeiro, G. & Williams, R. S. Observation of two resistance switching modes in TiO2 memristive devices electroformed at low current. *Nanotechnology* **22**, 254007/1-7 (2011).

28. Choi, S., Balatti, S., Nardi, F. & Ielmini, D. Size-dependent drift of resistance due to surface defect relaxation in conductive-bridge memory. *IEEE Electron Device Lett.* **33**, 1189-91 (2012).

29. Mitkova, M., Kozicki, M. N., Kim, H. C. & Alford, T. L. Thermal and photodiffusion of Ag in S-rich Ge-S amorphous films. *Thin Solid Films* **449**, 248-253 (2004).

30. Cho, D.-Y., Valov, I., van den Hurk, J., Tappertzhofen, S. & Waser, R. Direct Observation of Charge Transfer in Solid Electrolyte for Electrochemical Metallization Memory. *Advanced Materials* **24**, 4552-4556 (2012).

31. Chua, L., Sbitnev, V. & Kim, H. Hodgkin–Huxley Axon is Made of Memristors. *International Journal of Bifurcation and Chaos* **22**, 1230011/1-48 (2012).

32. Menzel, S., Böttger, U. & Waser, R. Simulation of multilevel switching in electrochemical metallization memory cells. *J. Appl. Phys.* **111**, 014501/1-5 (2012).

33. ITRS, The International Technology Roadmap for Semiconductors - ITRS 2011 Edition. (2011).

34. Rosezin, R. *et al.* Verfahren zum nichtdestruktiven Auslesen resistiver Speicherelemente und Speicherelement. *German Patent DE102011012738* (2012).

35. Meuffels, P. & Soni, R. Fundamental Issues and Problems in the Realization of Memristors. *arXiv:1207.7319 [cond-mat.mes-hall]*, 1-14 (2012).

36. Saraf, S., Markovich, M., Vincent, T., Rechter, R. & Rothschild, A. Memory diodes with nonzero crossing. *Appl. Phys. Lett.* **102**, 22902/1-4 (2013).

37. van den Hurk, J., Valov, I. & Waser, R. Preparation and characterization of $GeS_x$ thin-films for resistive switching memories. *Thin Solid Films* **527**, 299-302 (2012).

38. Bard, A. & Faulkner, L. *Electrochemical Methods: Fundamentals and Applications* (John Wiley and Sons, New York, 2001).

39. Di Ventra, M., Pershin, Y. V. & Chua, L. O. Circuit Elements With Memory: Memristors, Memcapacitors, and Meminductors. *Proc. IEEE* **97**, 1717-1724 (2009).

## Author contributions

I.V. conceived the idea, designed the experiments, interpreted the data, wrote the manuscript; E.L. performed the memristive simulations, co-wrote the manuscript, contributed to data interpretation; S.T. prepared the ECM cells, performed the measurements, contributed to data interpretation; S.S. performed measurements on VCM cells; J.v.d.H. prepared $GeS_x$ cells; F.L. prepared the $WO_x$ cells and performed measurements; R.W. initiated and supervised the research, contributed to the concept of the study. All authors discussed the results and implications at all stages and contributed to the improvement of the manuscript text.

## Acknowledgements

We gratefully acknowledge financial support in parts by the Deutsche Forschungsgemeinschaft through the SFB 917 “Nanoswitches” and the project WA 908/22, as well as by Intel Corp and by Samsung Advanced Institute of Technology (SAIT – GRO Program). We would like to thank S. Menzel for fruitful discussions on the kinetics of the resistive switching, P. Meuffels for checking our memristor equations, and R. Meyer for efficient proof-reading and excellent comments.

## Supplementary Information

### S1 On the use of the term 'Memristor'

The memristor was theoretically derived in 1971 by L. Chua as a non-linear circuit element correlating flux-linkage $\phi$ and charge $q$ [1]. From an axiomatic point of view, this device was considered the fourth element besides non-linear resistors, non-linear inductors and non-linear capacitors. The hypothetical device offers a one-to-one mapping of $\phi$ and $q$ and can be either described as a voltage or current controlled device.

Voltage controlled:

$$I = G(\phi) \cdot V$$
$$\dot{\phi} = V$$

Current controlled:

$$V = R(q) \cdot I$$
$$\dot{q} = I$$

In 1976 the memristor concept was generalized towards memristive systems by L. Chua and S. M. Kang to cover a wide range of real devices [2]. Corresponding memristive elements are characterized by an *I-V* hysteresis loop pinched at the origin, which the authors call a *zero-crossing property.* A memristive element offers an arbitrary number of internal state-variables $\mathbf{x} = \left[x_1 ... x_n\right]$.

Voltage controlled:

$$I = G(x, V) \cdot V$$
$$\dot{\mathbf{x}} = f(\mathbf{x}, V)$$

Current controlled:

$$V = R(x, I) \cdot I$$
$$\dot{\mathbf{x}} = f(\mathbf{x}, I)$$

Several physical devices such as thermistors and discharge tubes were modeled as memristive elements in ref. [2]. Moreover, a Hodgkin-Huxley neuron could be modeled as memristive element and an additional battery in series. Note that this modeling approach is not feasible for ReRAM cells where the controlling ionic battery path is in parallel to the controlled electronic resistance (see S10).

In 2008 $TiO_x$ based VCM devices were identified as memristive elements by Strukov et al. [3], In 2011 L. Chua suggested to use the term 'memristor' for any ReRAM cell [4] and calling any memristive systems a 'memristor'. Please note this 'real' memristor does not represent the fourth element anymore and must not be confused with the (still hypothetical) 'ideal' memristor (definition from 1971 [1]). In the line of the redefinition ReRAM cells have been considered 'memristors' [4]. In our paper we show that the redefinition may be true for e.g. STT-MRAM devices, however, this is not applicable for ReRAM cells because of the inherent nanobattery which violates the *I-V* pinched zero-crossing properties. For this reason, we suggest an extension of the memristor theory as outlined in the main text.

## S2 Contribution of $V_{GT}$

The contribution of $V_{GT}$ in accordance to equation (6) is presented in Fig. S2.1. We programmed three different ON states in a $Ag/GeS_{2.2}/Pt$ system, characterized by three different resistances of 600 Ω, 1500 Ω and 3500 Ω, by setting two different current compliances of 100 µA, 10 µA and 1 µA, respectively. Please note that the devices we used for this experiment were deliberately not equipped with a thin oxide barrier in the electrolyte, so that the ON states were more unstable than in usual devices. In devices targeting memory applications such diffusion barriers are typically introduced to prevent a fast chemical dissolution of Ag into the glassy chalcogenides matrix and to limit the ON current densities. In Fig. S2.1 it is shown that all emfs in the ON state exhibit initially a short circuit resistance. The emf value (and the ON resistance) of the cell programmed at 100 µA remains constant within a long period of time indicating that the metallic filament was stable. For the cells programmed at current compliances of 10 µA and 1µA after a limited period of time a sharp increase of the emf was observed. This increase obviously corresponds to dissolution of the nanosized filament. In accordance to equation (6) the voltage is of a positive polarity. However, the impact of $V_{GT}$ cannot be separated by the impact of the diffusion emf i.e., $V_d$ (which is also of positive polarity). The voltage equilibrates at values typical for the OFF state within 60 seconds. The subsequently measured high resistances confirm that the cells indeed underwent a RESET to OFF state. Similar behavior was also observed for $WO_{3-x}$ and $GeSe_{2.3}$ based cells.

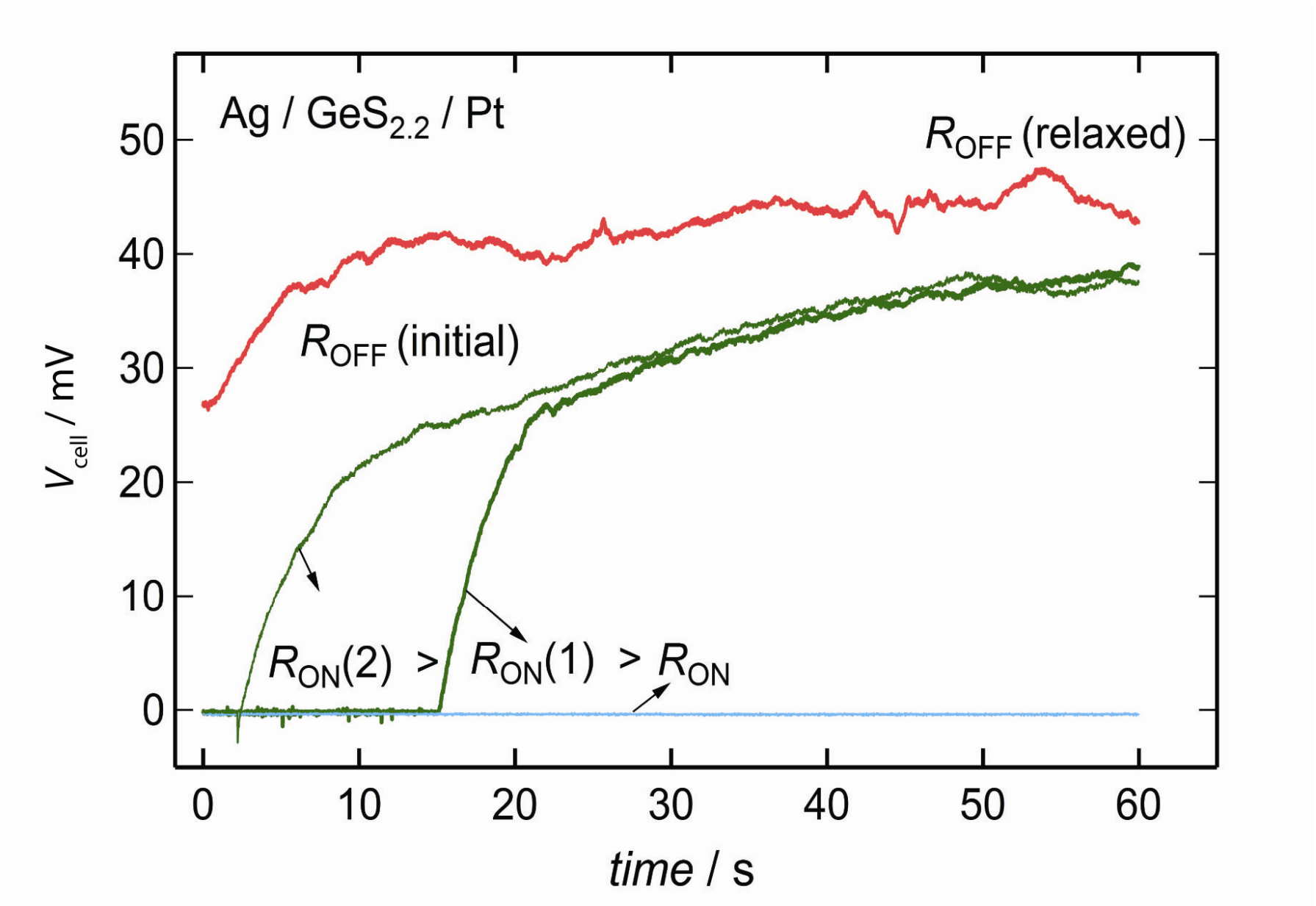

**Figure S2.1** Steady state emfs for the OFF state and the RESET transition from ON to OFF state corresponding to a dissolution of the filament. The dissolution is induced by three different factors: $V_{GT}$, $V_d$, and also chemical reactions of dissolution within the electrolyte matrix[5].

## S3 Measurement Methods

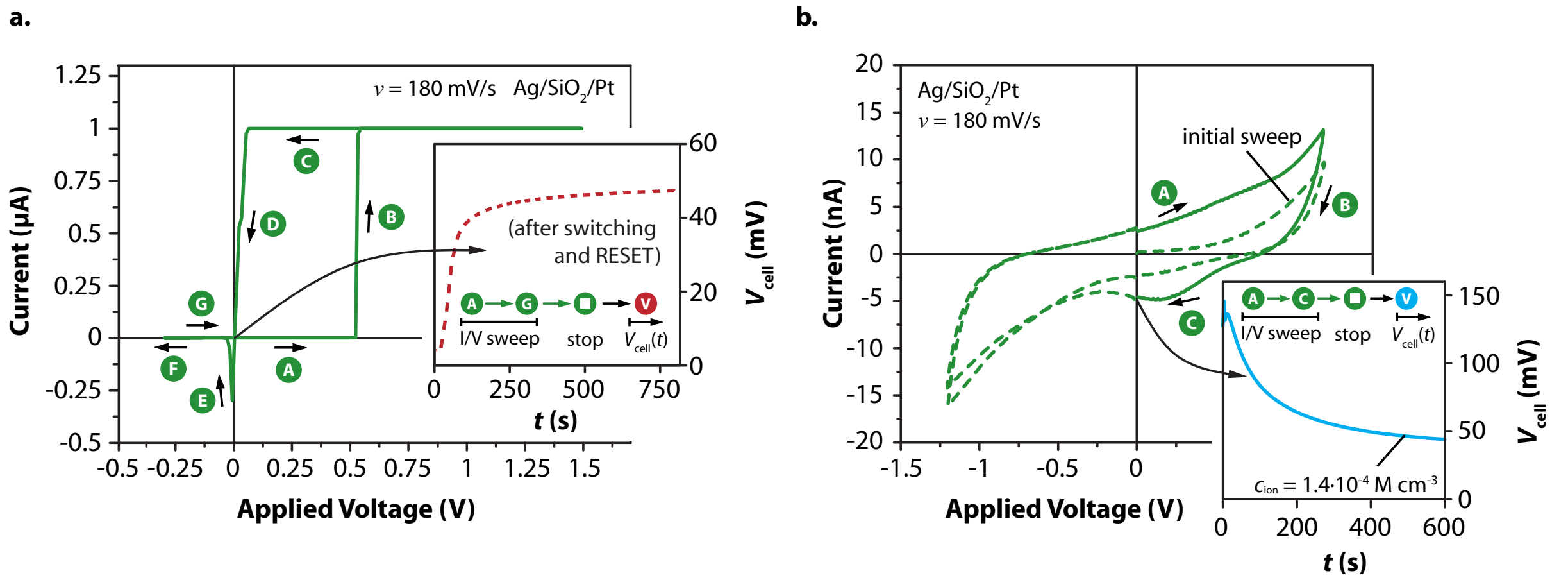

**Figure S3.1** Current voltage characteristics for oxidation of Ag and reduction of $Ag^+$ ions during device operation of a $Ag/SiO_2/Pt$ based ECM cell. The transient emf measurements correspond to Fig. 2b,c **(a)** Resistive Switching of a previously formed $Ag/SiO_2/Pt$ cell. For readability the switching curve is labeled by (A) to (G). Initially, the cell was SET into the ON state (A) to (D) and subsequently RESET into the OFF state (E) to (G). After RESET the *I*/*V* sweep was stopped and the emf was immediately measured (red dotted line). **(b)** Cyclic voltammetry of a similar $Ag/SiO_2/Pt$ cell as used for resistive switching. Without forming[6] and by limitation of the voltage amplitude (-1.5 V $\leq V \leq$ 1.5 V) the $Ag^+$/Ag redox-process can be observed[7]. The ion concentration $c_{ion}$ can be adjusted by the sweep rate during oxidation and $V > 0$ (A) to (C).[8] The emf as a function of the ion concentration was subsequently measured. Details on cyclic voltammetry can be found in[9].

**S4 Derivation of equation (10) on the basis of linear combination of equations (5) and (9)**

The potential determining electrode reaction at the interface $Ag/SiO_2$ (s'):

$$Ag^+ + e^- \rightleftharpoons Ag \quad \text{with} \quad K_{Ag} = \frac{a_{Ag}}{a_{Ag^+} a_{e^-}} \tag{s1}$$

The potential determining electrode reaction at the interface $Pt/SiO_2$ (s"):

$$\tfrac{1}{2}O_2 + H_2O + 2e^- \rightleftharpoons 2OH^- \quad \text{with} \quad K_{H_2O} = \frac{a_{OH^-}^2}{a_{O_2}^{1/2} \cdot a_{H_2O} \cdot a_{e^-}^2} \tag{s2}$$

Nernst potential contribution:

$$V_N = V^0 + \frac{kT}{2e} \ln \frac{(a_{Me^{z+}}^2)_{s'} \cdot (a_{OH^-}^2)_{s"}}{(a_{Me}^2)_{s'} \cdot (a_{O_2}^{1/2})_{s"} \cdot (a_{H_2O})_{s"}} \tag{s3}$$

Diffusion potential contribution ($z_{Me^+} = +1$ for $Ag^+$ and $z_- = -1$ for electrons and $OH^-$):

$$V_d = -\frac{kT}{e} \sum_i \int_{s"}^{s'} \frac{t_i}{z_i} d \ln a_i = -\frac{kT}{e} \left( \frac{\bar{t}_{Me^+}}{z_{Me^+}} \ln \frac{(a_{Me^+})_{s'}}{(a_{Me^+})_{s"}} - \frac{\bar{t}_-}{z_-} \ln \frac{(a_-)_{s'}}{(a_-)_{s"}} \right)$$

$$= -\bar{t}_{Ag^+} \frac{kT}{e} \ln \frac{(a_{Ag^+})_{s'}}{(a_{Ag^+})_{s"}} - \bar{t}_{OH^-} \frac{kT}{e} \ln \frac{(a_{OH^-})_{s'}}{(a_{OH^-})_{s"}} - \bar{t}_{e^-} \frac{kT}{e} \ln \frac{(a_{e^-})_{s'}}{(a_{e^-})_{s"}} \tag{s4}$$

In accordance to equations (s1) and (s2) one substitutes:

$$(a_{e^-})_{s'} = \frac{(a_{Ag})_{s'}}{(a_{Ag^+})_{s'} K_{Ag}}, \quad (a_{e^-})_{s"} = \frac{(a_{OH^-})_{s''}}{(a_{O_2}^{1/4})_{s''} \cdot (a_{H_2O}^{1/2})_{s''} \cdot K_{H_2O}^{1/2}} \quad \text{and} \quad \bar{t}_{e^-} = 1 - \bar{t}_{OH^-} - \bar{t}_{Ag^+}.$$

In this way the term of electronic transference number is expressed as:

$$\bar{t}_{e^-} \ln \frac{(a_{e^-})_{s'}}{(a_{e^-})_{s"}} = \left(1 - \bar{t}_{OH^-} - \bar{t}_{Ag^+}\right) \ln \frac{(a_{Ag})_{s'} (a_{O_2}^{1/4})_{s''} \cdot (a_{H_2O}^{1/2})_{s''} \cdot K_{H_2O}^{1/2}}{(a_{Ag^+})_{s'} K_{Ag} (a_{OH^-})_{s''}} \tag{s5}$$

Or equation (s4) reads:

$$V_d = -\bar{t}_{Ag^+} \frac{kT}{e} \ln \frac{(a_{Ag^+})_{s'}}{(a_{Ag^+})_{s"}} - \bar{t}_{OH^-} \frac{kT}{e} \ln \frac{(a_{OH^-})_{s'}}{(a_{OH^-})_{s"}}$$
$$- \left(1 - \bar{t}_{OH^-} - \bar{t}_{Ag^+}\right) \frac{kT}{e} \ln \frac{(a_{Ag})_{s'} (a_{O_2}^{1/4})_{s''} \cdot (a_{H_2O}^{1/2})_{s''} \cdot K_{H_2O}^{1/2}}{(a_{Ag^+})_{s'} K_{Ag} (a_{OH^-})_{s''}} \tag{s6}$$

The emf of the cell is determined by the sum of the both contributions:

$$V_{\text{emf}} = V_{\text{N}} + V_{\text{d}} \tag{s7}$$

Substituting equation (s3) and equation (s6) in equation (s7) and accounting for the canceling and constant terms in the logarithms one derives the equation:

$$V_{\text{emf}} = \underbrace{V^0 + const.}_{V_0} + \bar{t}_{\text{OH}^-} \frac{kT}{e} \ln(a_{\text{Me}^+})_{s'} + \bar{t}_{\text{Me}^+} \frac{kT}{e} \ln(a_{\text{OH}^-})_{s''} \sim V_0 + \bar{t}_{\text{ion}} \frac{kT}{2e} \ln(a_{\text{ion}}) \tag{s8}$$

The emf in accordance to eq. (s7) and respectively, eq. (s8) represents the general case of possible origins of the emf. In fact not all contributions must always be present. For example the diffusion potential contribution (eq. (s4)) can be minimized or even completely eliminated in case of charged species with a high mobility, i.e. fast relaxation. Nernst potential (eq. (s3)) can be eliminated by using the same electrode materials. However, a complete simultaneous elimination of non-equilibrium conditions for ReRAM devices of practically application is not possible. Thus, the emf is perpetually induced in these devices during the operation and/or reading cycles but it can relax with time. The relaxation time depends strongly on the particular system, the transport properties of the conducting solid and the thermodynamic conditions.

## S5 Reduction Current and Short Circuit Measurement

Fig. S5.1 depicts the situation when the cell was first SET and then RESET. The performed continuous subsequent cathodic cycles showed that the current further decreases with each cycle. This behavior clearly confirms that the measured non-pinched characteristics are due to the emf and not due to dielectric capacitance effects.

An additional confirmation is provided by a cyclic voltammetry study (*I-V* sweeps). The measured currents at a given potential value are linearly dependent on the sweep rate in the case of dielectric capacitive effects and proportional to a square root of the sweep rate in the case of Faraday currents, respectively. In our previous papers[7, 8] we demonstrated that in Cu/$SiO_2$/Pt systems the condition of a linear relation $I$ vs. $V^{1/2}$ is fulfilled. The observed Faraday peaks in the cyclic voltammogram also unequivocally proved that the currents are due to electrochemical reactions[8]. The emf in the ReRAM cells acts as a nano-battery as confirmed by the typical characteristics for a battery discharge shown in Fig. S5.2 for $SiO_2$ based ECM cells with various electrode diameter $d$ (electrode area $A_{\text{cell}}$).

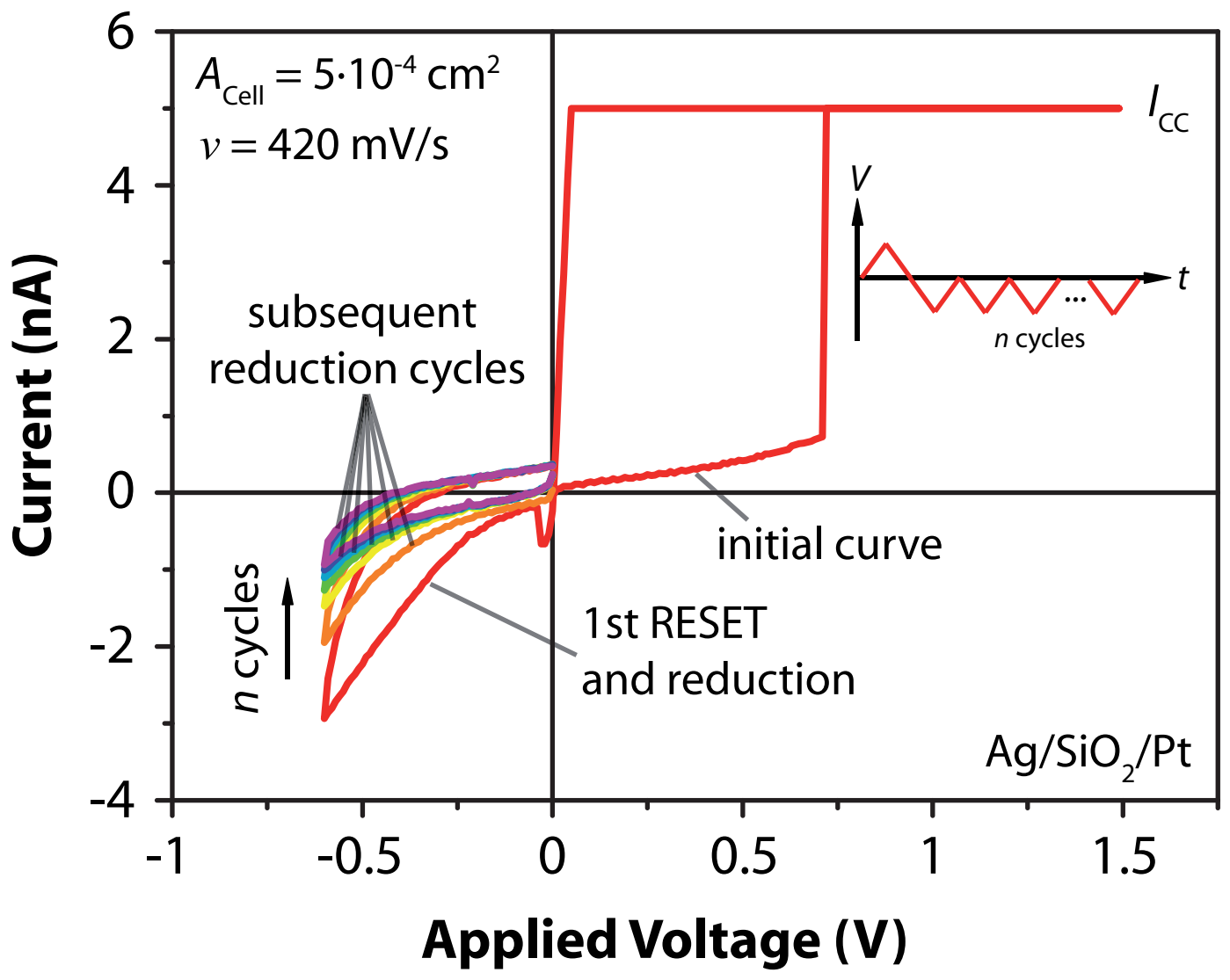

**Figure S5.1** Resistive Switching of a Ag/SiO$_2$/Pt cell using a current compliance $I_{CC}$ = 5 nA. The cell was initially SET and RESET afterwards. After the first RESET subsequent current/voltage sweeps were performed in the negative voltage regime. The current clearly decreases by further cycling which is contributed to further reduction of previously oxidized silver ions. This behavior cannot be attributed to a trivial polarization effect. The non-pinched hysteresis is clearly demonstrated.

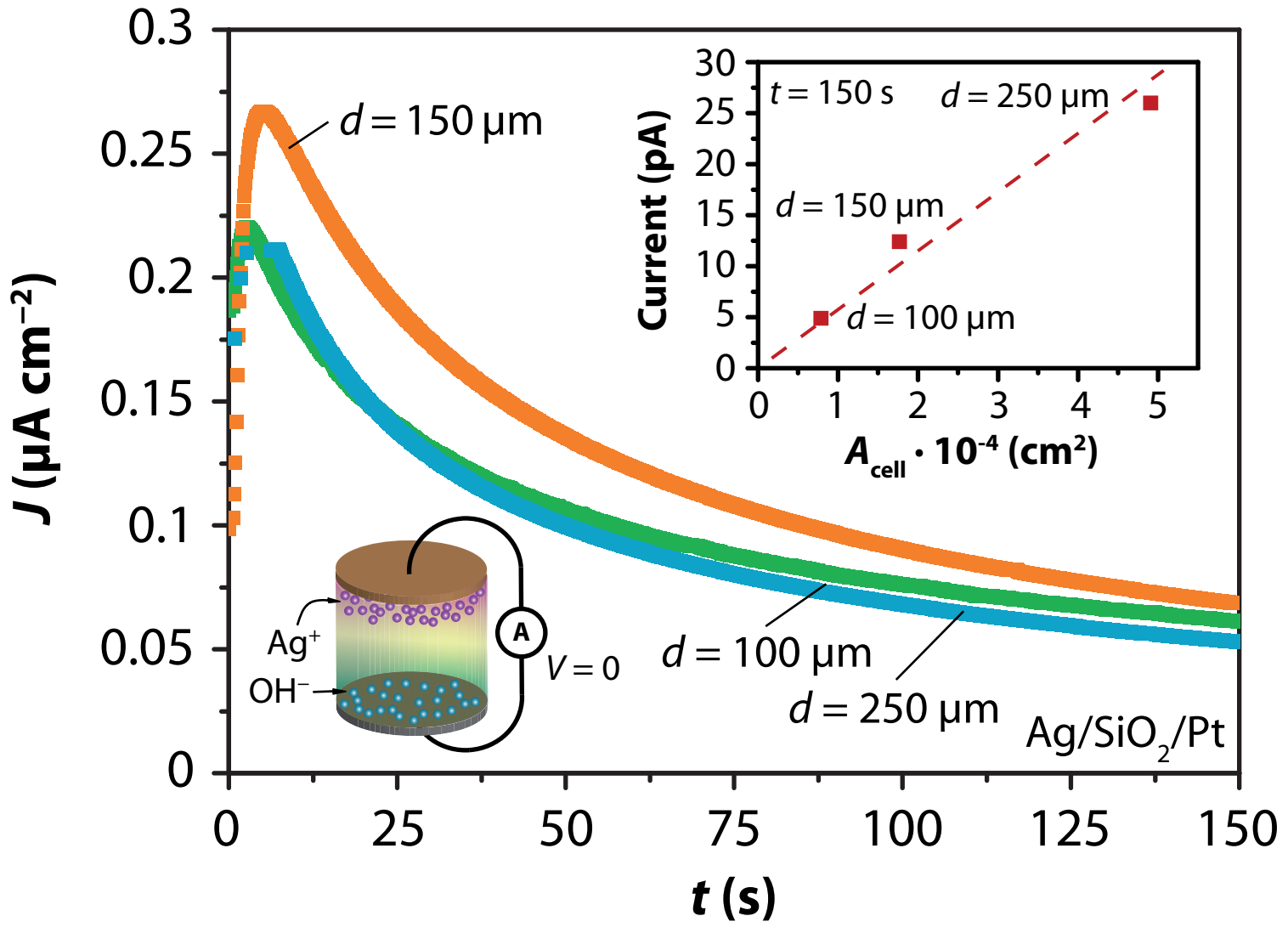

**Figure S5.2** Short circuit ($V$ = 0) current density $J$ measurement of a previously switched Ag/SiO$_2$/Pt cell after RESET. A significant short circuit current is observed for several seconds and depends on the electrode diameter $d$. The short circuit current cannot be fitted to a discharge of an electronically polarized capacitance. The inset depicts the cell current at $t$ = 150 s for $d$ = 100 µm, $d$ = 150 µm and $d$ = 250 µm. The measurement resolution is smaller than 0.01 pA.

## S6 Emf and volatility of the OFF state

The nanobattery effect is observed mainly for the OFF state. There is self-discharge (i. e. current flow under open-circuit conditions) of the nanobattery over $R_i$ and $R_e$ (Fig. 2a), where $R_e$ may be mainly determined by $R_{leak}$ in the OFF state (Fig. 5). This self-discharge process typically shows at least two time constants where the first time constant is in the order of some hundred seconds (see e.g. Fig. 2c) and a second time constant is in the order of many thousand seconds. For the VCM-type cells we typically observe a faster self-discharge obviously due to the lower $R_e$. In any case, the self-discharge of these ReRAMs will occur at times much shorter than 10 years (retention time requirement for non-volatile memories). However, the relaxation of the voltage does not lead to a loss of the logic state (which would mean that the cell changes from the initial OFF state into the ON state due to the self-discharge). However, the relaxation may lead to a certain drift of the OFF state resistance. Of this reason it is important to account for the existence of the nanobattery effect in the ReRAM cells. For example, the OFF resistance of nanoscale $TiO_x$ based cells drifts from 100 MΩ to 1-100 GΩ within 3 weeks as reported in Ref. [10]. Shifts in $R_{OFF}$ are also reported for ECM cells e.g. Ref. [11] These drift processes may easily be caused by an emf voltage and would then demonstrate the relevance to device retention issues.

Thus, it is important to acknowledge that an emf is not necessarily leading to volatility of the state. [12]

## S7 Downscaling ReRAM devices

For pristine cells, properties such as the leakage current are area proportional within the first approximation. However, after electroforming this is not true anymore for filamentary ReRAM types. The ON state is determined by a metallic filament for ECM cells and a highly electronic-ionic conducting filament for VCM cells, leading to $R_{ON}$ values independent of the area. Also in the OFF state, there is a fast ion transport channel along the original filament for ECM cells (this is the reason for the much lower SET voltage compared to the forming voltage). In the OFF state of VCM cells, most of the highly electronic-ionic conducting filament remains (this stub of the filament is called "plug"). The filament is just disrupted by a thin oxidized and, hence, more insulating barrier ("disc") close to the active electrode (Ref. [13]. For this reason, a downscaling from micrometer sized cells to nanometer sized cells will lead to an increase in the dominance of the conductance of the filament compared to the rest of the area (Fig. S7.1).

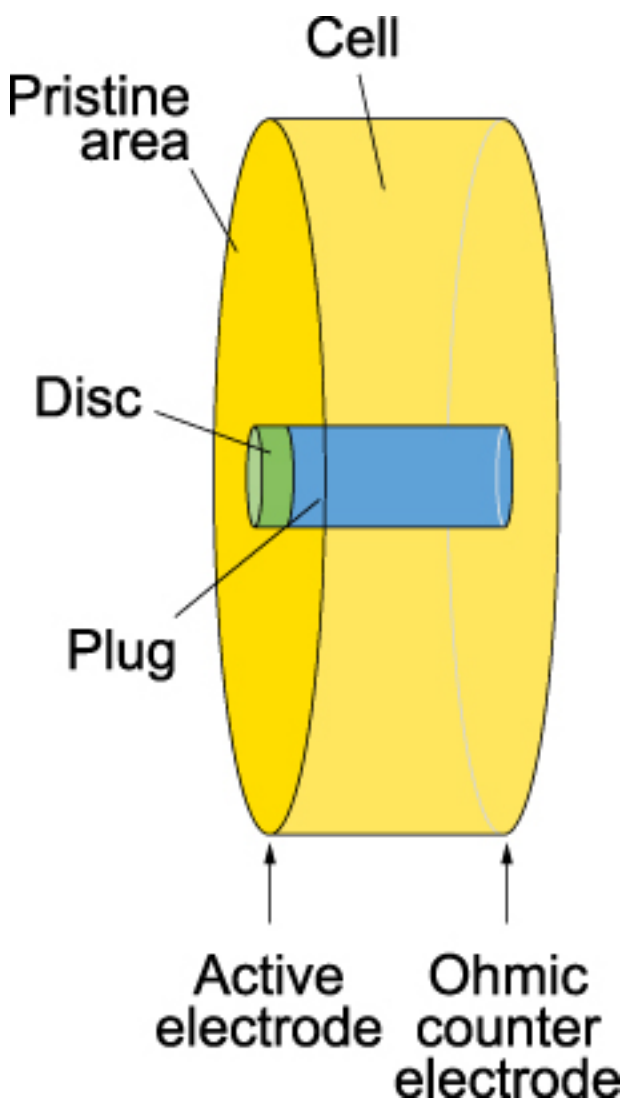

**Fig. S7.1** Sketch of the resistance components which make up the resistance of a VCM cell. In parallel to the (low) resistance of the filament, there is the resistance of the of the unformed, pristine film. The ON and OFF state are controlled by a higher or lower resistance of the "disc" at the end of the filament close the active electrode (left), while the stub ("plug") of the filament remains unchanged in first approximation (reproduced from [13]).

A $V_{cell}$ measurement for nanoscale VCM type TiN/$TiO_2$/Ti cells is shown in Fig. S7.2. Please note that this are preliminary results which awaits confirmation by extended statistical studies. The result indicates that indeed $V_{cell}$ can be detected for nanoscale ReRAM cells.

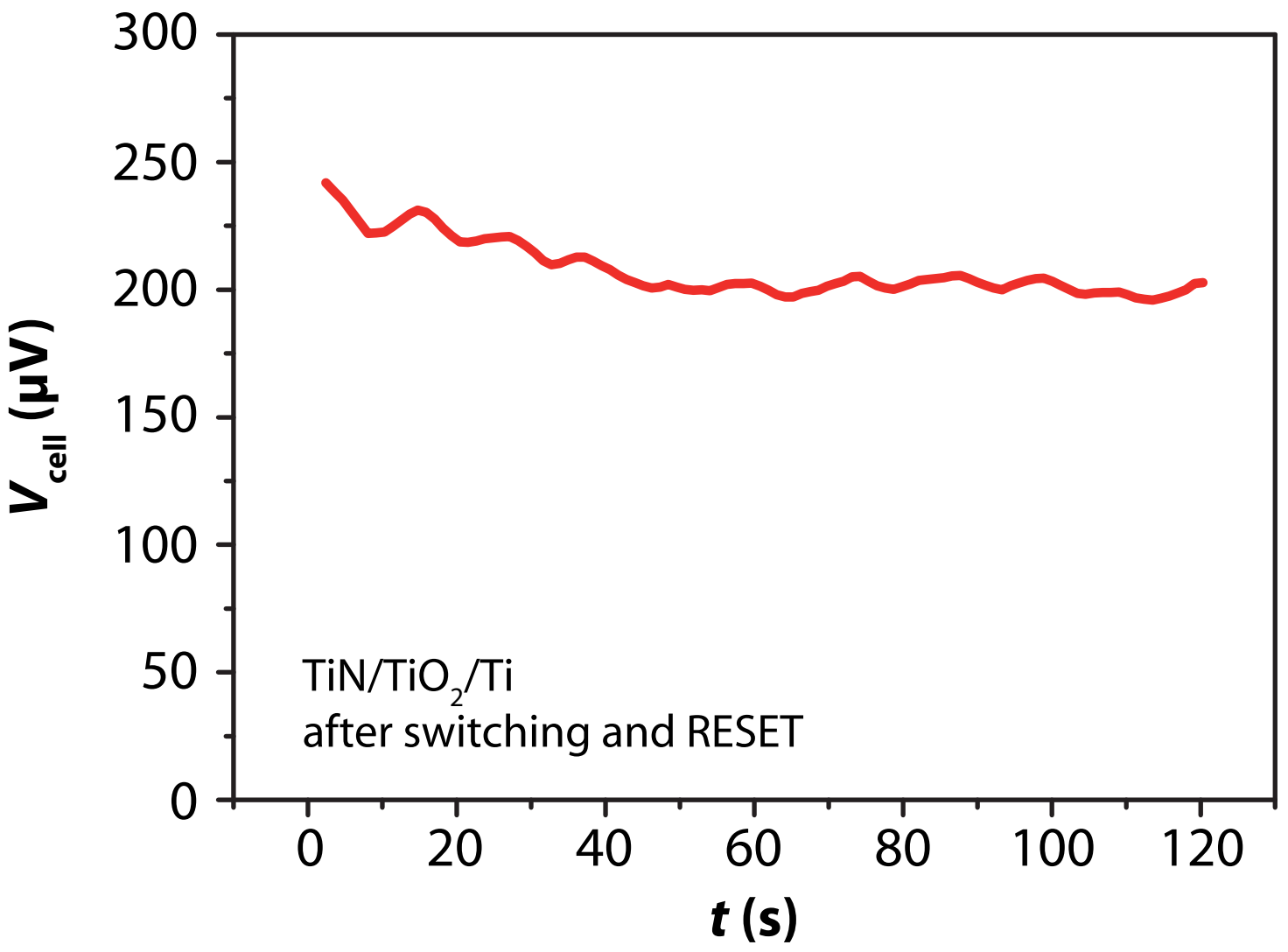

**Fig. S7.2** Transient cell voltage measurement of a Ti/$TiO_2$/TiN nanocrossbar cell (≈ 55 nm · 55 nm). The measurement has been immediately performed after switching and RESET of the cell. The 10 nm thick $TiO_2$ layer has been fabricated by reactive sputtering. A Pt layer has been additionally deposited on the Ti top electrode In order to prevent chemical oxidation of Ti in air.

In general, if high-temperature bulk diffusion data are used to estimate the switching kinetics of ReRAMs, great care must be exercised. For instance, the Ag cation diffusion coefficients in $SiO_2$ [14] measured at temperatures between 525 K and 875 K and extrapolated to slabs of 10 nm thickness at 300 K lead to a diffusion coefficient of $2.10^{-23}$ $cm^2$/s (electrical mobility of ~ $4.10^{-22}$ $cm^2$/Vs). Hence, switching of a ECM-type Ag/$SiO_2$/Pt cell (as used in our study) at e.g. 1 V would need over 75 years (!) to switch. In VCM cells similar calculations (based on the paper by Noman et al.[15]) lead to switching times (at $V$ = 1 V and film thickness of 10 nm) of more than 30 hours including a field acceleration component. In reality, however, we (and many other groups working in this field) can switch such cells in < 10 ns. Translated into $R_i$ (assuming the same concentration of mobile charges), there is > 10 orders of magnitude difference between the extrapolated bulk values and the experimental thin film switching times. The reason for the discrepancy is not known yet. It may be speculated that the thin films deposited by sputtering, ALD, evaporation etc., typically at much lower temperature than bulk sintering or melting, show a nanoporosity which strongly enhances the ion mobility. Furthermore, it is conceivable that different diffusion paths (such as extended defects) start to dominate at lower temperatures. A significant acceleration (up to orders of magnitude) will arise for the field acceleration of the ion mobility due to the high fields (> 1 MV/cm) in the disc in front of the active electrode in assumed to be the range from 3 to 5 nm, responsible of the actual modification of the electrostatic barrier.

In principle, different diffusion regimes (e.g. due to approaching nano-dimensions [16, 17] and when the film thickness is comparable to the space charge layer thickness [18])[19] are well known in material science for many years. In VCM cells, a disc in front of the active electrode which is smaller in thickness than typically space charge layers [20, 12].

## S8 Steady state cell voltages for ECM (Cu/$SiO_2$/Pt) and VCM (Pt/$SrTiO_3$/Ti) cells

In the as-deposited state the ReRAM cells based on the valence change mechanism (VCM) show a cell voltage which is comparable to that of the ECM cells. Fig. S8.1 shows the steady state cell voltages of as-deposited (unformed) Pt/$SrTiO_3$/Ti VCM and Cu/$SiO_2$/Pt ECM cells. Both VCM and ECM systems show cell voltages in the order of some hundreds of millivolts.

However, after the forming process the cell voltage drops from the range of millivolts down to microvolts for VCM cells and the cell resistance drops by some orders of magnitude ($R_{pris}$ ~ $10^4$ $R_{OFF}$ ). As shown in Fig. S8.2, the cell voltage takes values in the range of some millivolts (STO system) to tens of microvolts ($Ta_2O_5$ system). These much lower cell voltages values in the formed OFF state might at a first glance indicate lower chemical potential gradients in these cells. However, it is more likely caused by the fact that during the formation process the oxide gets partially reduced which gives rise to an increase of the electronic partial conductivity, leading to the drop in the cell resistance. This means that the transference number of the electrons $t_e$ becomes much higher, i.e. much closer to 1. Because of $t_{ion} + t_e = 1$, this means that $t_{ion}$ (here the oxygen ion transference number $t_O$) drops to a very small value of approximately $10^{-3}$ (or even below). The measured cell voltage decreases correspondingly.

Note that this interpretation is consistent with results for nano-structured VCM type devices (see S7). Additional results on nanocrossbar cells will be published in a forthcoming paper.

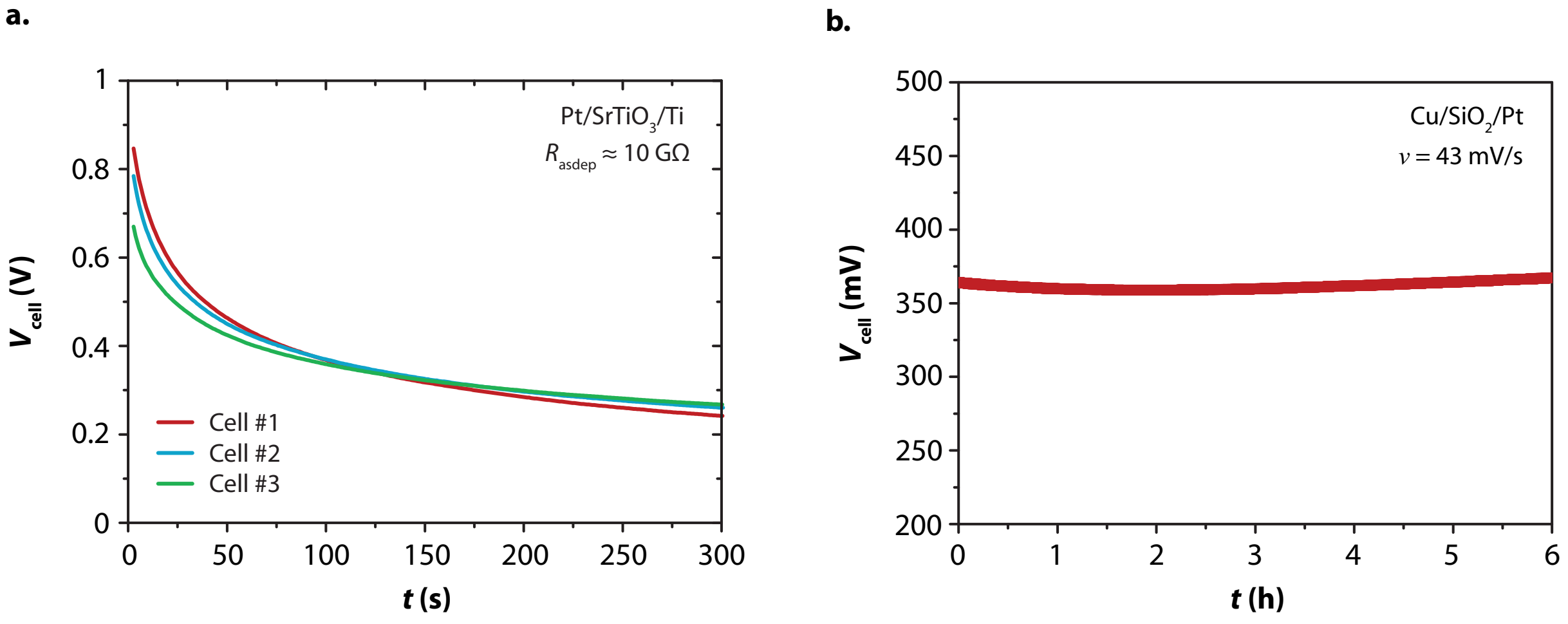

**Figure S8.1.** Steady state cell voltages for as-deposited cells after anodic oxidation without forming and resistive switching **(a)** Pt/$SrTiO_3$/Ti and **(b)** Cu/$SiO_2$/Pt.

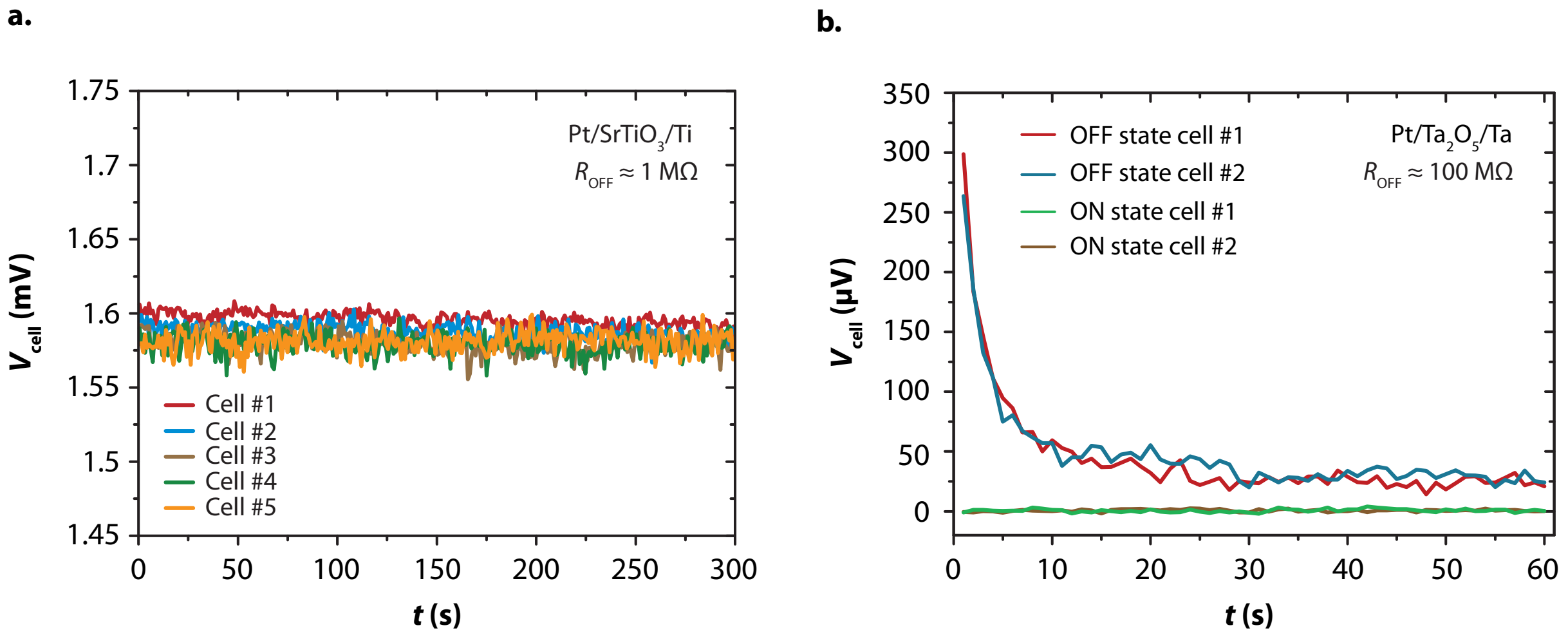

**Figure S8.2** Steady state cell voltages for **(a)** Pt/$SrTiO_3$/Ti and **(b)** Pt/$Ta_2O_5$/Ta based VCM cells after switching in the OFF state. In addition, the cell voltage in the ON state is shown in (b) which can be clearly distinguished from the cell voltages in the OFF state.

## S9 Non zero-crossing hysteresis of ECM and VCM cells

The current-voltage characteristics of Ag/$SiO_2$/Pt (Fig. S9.1) cells are similar to those shown in Fig. 4b for Cu. Clearly, the *I-V* crossing does not occur in the origin.

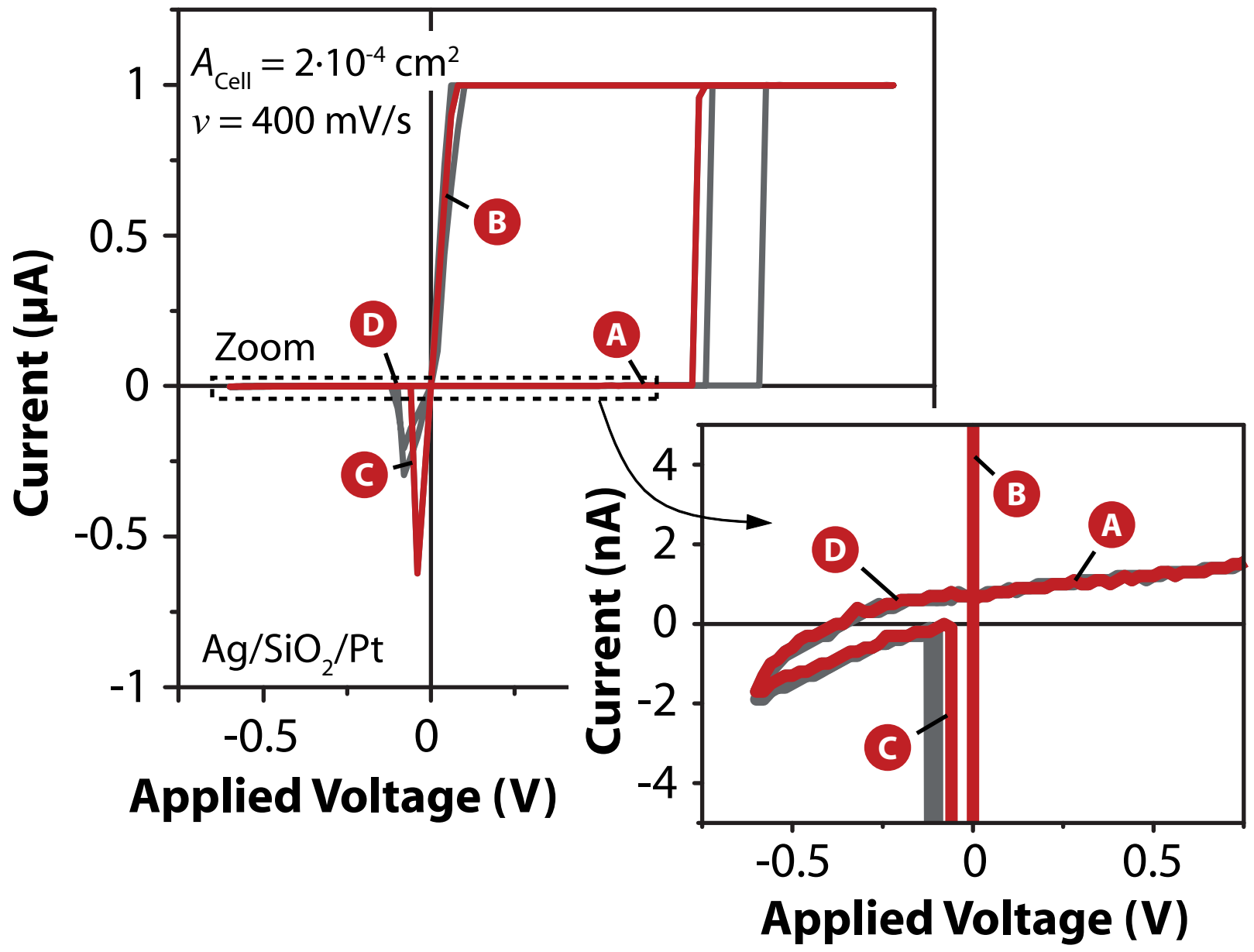

**Figure S9.1** Resistive Switching of a Ag/$SiO_2$/Pt cell comparable to the Cu based cell depicted in Fig. 4b. The red curve is highlighted while statistical variation of subsequent cycles is depicted in grey color. For readability the current is labeled by (A) to (D). The non zero-crossing characteristic after RESET is clearly observed in the inset.

Similar to the ECM cells the cyclic voltammetry with VCM cells is characterized by non zero-crossing characteristics with currents of up to 0.5 µA for the Ti/$SrTiO_3$/Pt system as shown in Fig. S9.2.

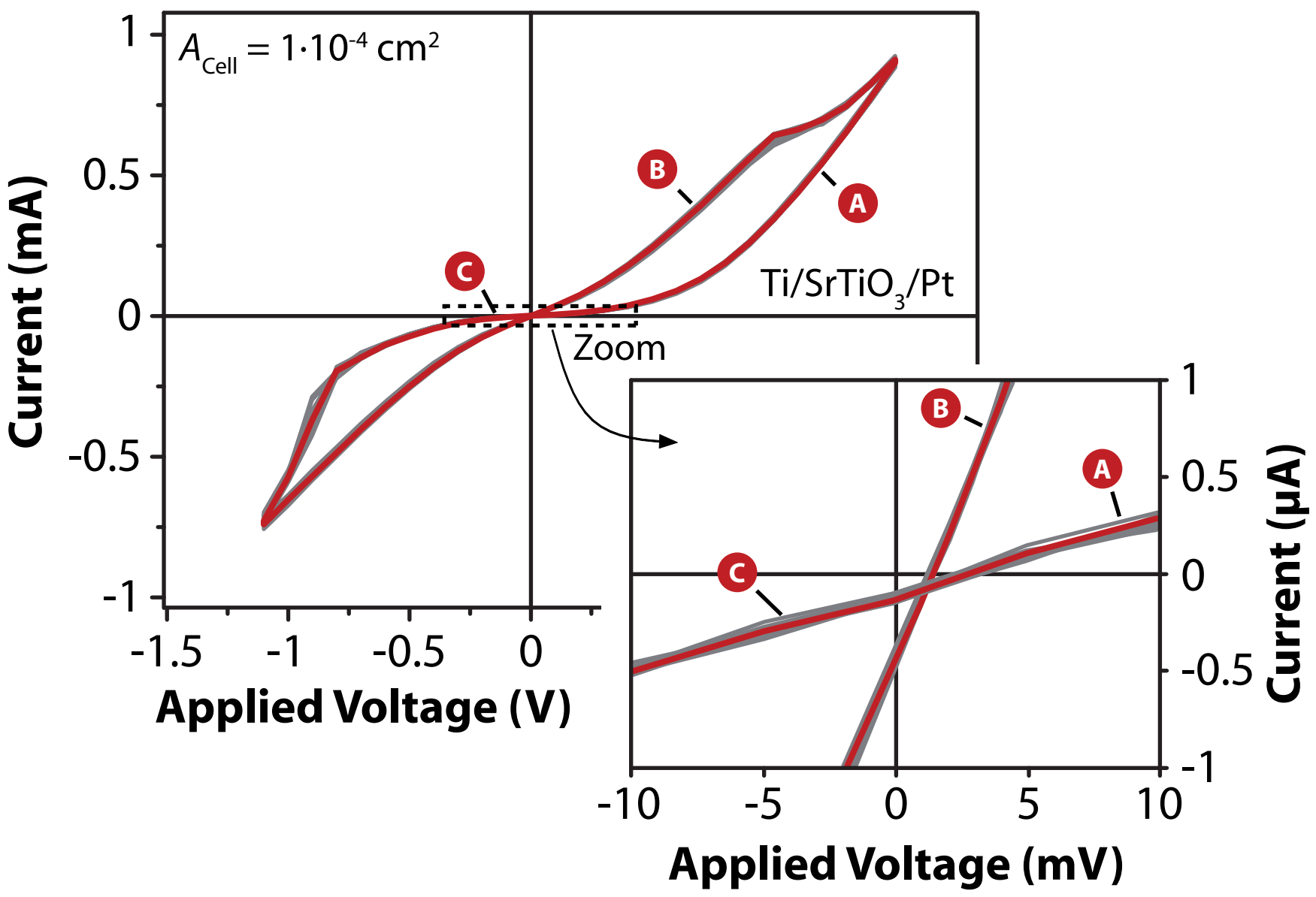

**Figure S9.2** Resistive Switching of a Ti/$SrTiO_3$/Pt cell. The red curve is highlighted while statistical variation of subsequent cycles is depicted in grey color. For readability the current is labeled by (A) to (D). The non zero-crossing characteristic after RESET is clearly observed in the inset.

We should note that in case of VCM cells filaments composed of a partially reduced phase can also facilitate a non-zero cell voltage in the ON state. As long as some minor part of the total conductivity remains ionic even the ON state can experience the cell voltage. This can be observed in the *I-V* sweep of Fig. S9.2 where an offset is also present for the ON state (and as expected smaller compared to the OFF state).

## S10 Simulation Details

*Extended memristive modelling*

Our extended memristive model contains three parallel current paths: The state-dependent electronic current path (filament), the state-independent electronic leakage current path (neglected for the simulation), and the ionic current path.

The resistance of the ionic current path ($R_i$) is a sum of the polarization resistances (Butler-Volmer equation) at both electrodes (anode $R_{p,a}$ and cathode $R_{p,c}$) and the ionic resistance $R_{ion}$ of the electrolyte (materials property) given by: $R_i(V) = R_{p,a} + R_{p,c} + R_{ion}$. $R_i$ is the derivative d$V$/d$I$ and can be influenced by each of the three components which all offer a highly nonlinear voltage dependence. The field acceleration of the ion mobility leads to pronounced decrease of $R_i$ for high field (above approx. 1 MV/cm) and, hence, to a strong non-linearity of $R_i$(V). Thus, the $R_i$ value depends on the materials system and the applied voltage and can be individually calculated for the particular conditions, i.e. kind of electrolyte, applied voltage, temperature etc. However, to ensure the feasibility of the ReRAM device operation $R_i(V)$ must have a finite value.

The two-dimensional state variable $\boldsymbol{x}$ consists of the structural variable (tunnelling gap) $x$ and the variable representing the ionic concentration $c_{ion}$:

$$\boldsymbol{x} = [x, c_{\text{ion}}] \qquad \text{(s9)}$$

The state-dependent Ohm's law of the extended memristive device reads:

$$I = I_{\text{el}}(V, x) + I_{\text{ion}}(V, c_{\text{ion}}), \qquad \text{(s10)}$$

and the corresponding two-dimensional state equation reads in case of anodic oxidation (i.e. charge and filament growth):

$$\dot{\boldsymbol{x}} = \begin{bmatrix} \dot{x} \\ \dot{c}_{\text{ion}} \end{bmatrix} = \mathrm{f}(\boldsymbol{x}, V) = \begin{bmatrix} K_1 \cdot \left| I_{\text{ion}}(V, c_{\text{ion}}) \right| \\ K_2 \cdot \left| I_{\text{ion}}(V, c_{\text{ion}}) \right| \end{bmatrix} \begin{matrix} \text{with } 0 \le x \le d \\ \text{with } c_{\min} \le c_{\text{ion}} \le c_{\max} \end{matrix} \qquad \text{(s11)}$$

The sign of $K_1$ and $K_2$ are reversed for cathodic sweep (discharging, i.e. $V < V_{emf}$).

To calculate the electronic current path, we apply the following tunnelling equation[21]:

$$I_{\mathrm{el}}(V,x)=\frac{eA_{\mathrm{fil}}}{2\pi hx^2}\left(\varphi_0-\frac{eV}{2}\right)\exp\left(-\frac{4\pi x}{h}\sqrt{2m_{\mathrm{eff}}}\sqrt{\varphi_0-\frac{eV}{2}}\right) \tag{s12}$$
$$-\frac{eA_{\mathrm{fil}}}{2\pi hx^2}\left(\varphi_0+\frac{eV}{2}\right)\exp\left(-\frac{4\pi x}{h}\sqrt{2m_{\mathrm{eff}}}\sqrt{\varphi_0+\frac{eV}{2}}\right).$$

Here, $A_{\mathrm{fil}}$ is the area of the filament ($r_{\mathrm{fil}}$ = 1 nm), $\varphi_0$= 3.6 eV is the barrier height, $m_{\mathrm{eff}} = m_0$ is the effective electron mass. The tunnelling equation accounts for the increasing electronic contribution shortly before switching to the ON state[22]. For the ionic current we use a Butler-Volmer equation to describe the electron transfer reaction of ions at the electrode interfaces::

$$I_{\mathrm{ion}}(V,c_{\mathrm{ion}})=I_0\cdot\sinh\left(\frac{V-V_{\mathrm{emf}}(c_{\mathrm{ion}})}{4kT/e}\right) \tag{s13}$$

For the simulation the following parameters were used (Ag/$SiO_2$/Pt cell):
$I_0$: exchange current (2 nA)
$\alpha$: charge transfer coefficient (0.5)
$z$: charge number (1)
$T$: temperature (300 K)

In equation (s11) and (s13), dynamic effects, in particular the impact of the sweep rate[7], diffusion limitation[8] and potential multi-step redox-reactions, are neglected to simplify the model. We assume identical interfaces (s' and s'') compare ref. [22].

The emf depends on the ion concentration, thus

$$V_{\mathrm{emf}}(c_{\mathrm{ion}})=V_0+\frac{kT}{2e}\ln\left(\frac{c_{\mathrm{ion}}}{c_0}\right) \tag{s14}$$

with $V_0$ = 0.17 V (see main text) holds true.

*Simulated I-V sweeps and influence of the dielectric capacitance*

In Fig. S10.1 the currents for the different sweeps do not overlay. Since the ion concentration during the sweeps changes, the emf varies correspondingly. The reason for changing the ion concentration is that during the cathodic sweep not all ions generated due to oxidation are reduced. The oxidation sweep continues up to 0.5 V whereas the cathodic one only to –0.3 V. Thus some ions remained and caused the current shift.

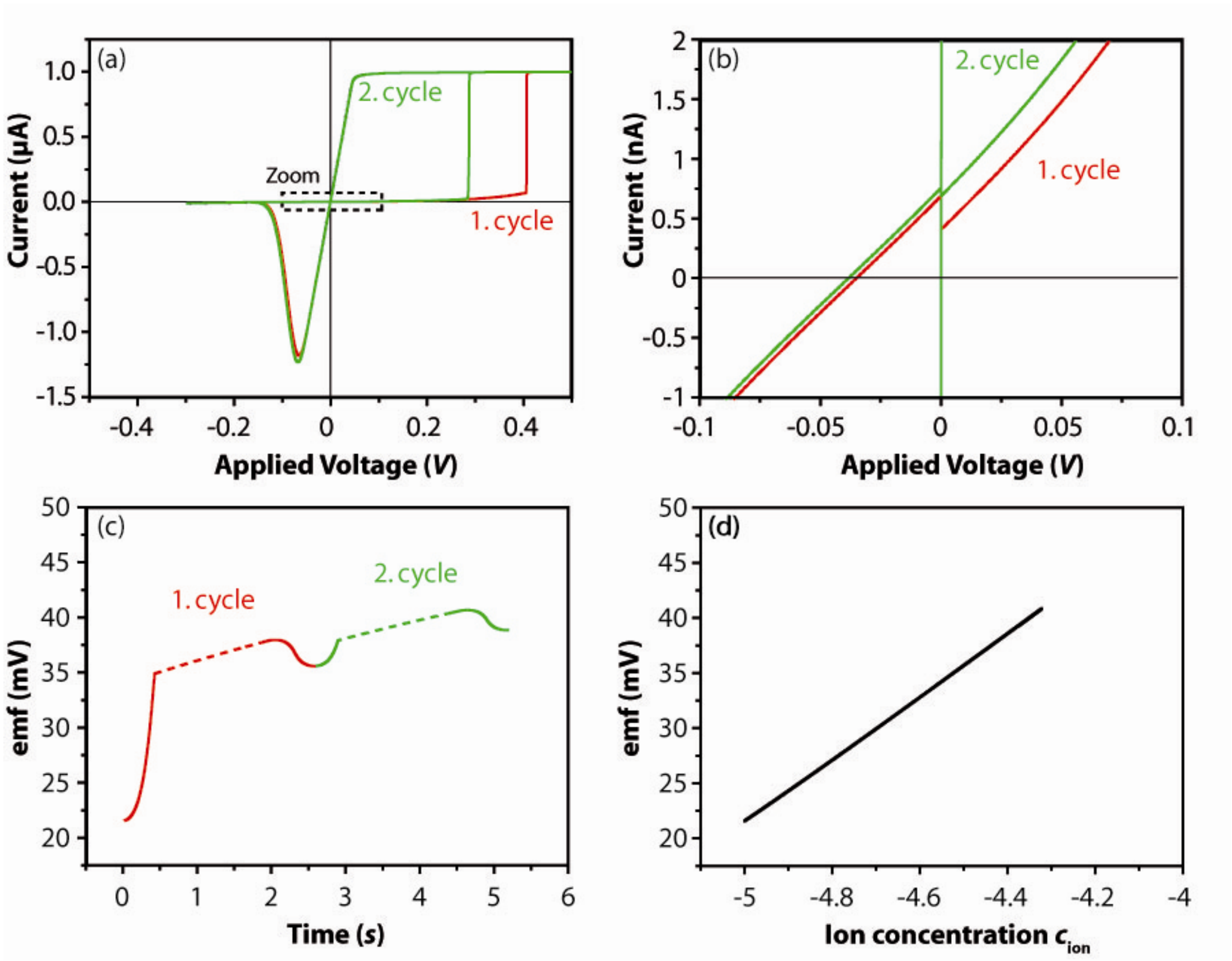

**Figure S10.1 (a)** First (marked by red color) and second cycle (marked by green color) of the extended memristive device simulation. **(b)** Zoom showing the non zero-crossing characteristic. The increase of the concentration $c_{\mathrm{ion}}$ leads to an increase of emf, hence the *y*-axis intercept varies from cycle to cycle. **(c)** $V_{\mathrm{emf}}$ (from (a)) as function of time. Note, $V_{\mathrm{emf}}$ cannot be experimentally determined in the ON state (dashed line) due to the electronic short circuit by the filament. **(d)** emf versus ion concentration.

The dielectric capacitance of the cells can be derived as follows:

$$C = \varepsilon_0 \varepsilon_{\mathrm{r}} \frac{A}{d} \tag{s15}$$

Here, the relative permittivity is assumed as $\varepsilon_{\mathrm{r}} = 5.5$ for $SiO_2$ [23], the area is $A = \pi\,(50\ \mu\mathrm{m})^2$ and the thickness is $d = 50$ nm. The corresponding capacitance for the Ag/$SiO_2$/Pt cell is $C = 7.6$ pF. In the simulation, a sweep rate of 1 V/s leads to a dielectric charging current of about 8 pA which is negligible compared to the current of 0.8 nA (Fig. S10.1b).

We checked a serial battery added to a memristive element as an alternative approach beside the model depicted in Fig. 5. The simulations show, however, that both ON and OFF states result in offsets on the voltage axis by the $V_{\mathrm{emf}}$. This is in clear contrast to our experimental findings showing a zero cell voltage for a short circuited (metallic) ON state and non-zero cell voltage for only the OFF state. Thus, a pure serial connection of a passive memristive element plus battery, as for example present in the Hodgkin–Huxley model[24], is not sufficient for ReRAM modelling.

### S11 Ionic resistances

The ionic resistance values $R_i(V)$ in accordance to the equivalent circuit model shown in Fig. 5b are provided along with the values for the total cell resistance ($R_{tot}$) and the electronic resistance ($R_e$) in Table S1 for all studied systems. We calculated these values on the basis of $V = V_{cell}$ and $I_{max}$ (at $V = 0$) or from the total resistance measured by impedance spectroscopy and the transference numbers $t_{ion} = (R_i/R_{tot})^{-1}$. Because $R_i$ depends on the surface area of the electrodes we provided the value $R_i/R_{tot}$ which is area independent. The $R_i/R_{tot}$ value is included in Fig. 4a.

| System | $R_{tot}$ | $R_i$ | $R_e$ | $R_i/R_{tot}$ |
|---|---|---|---|---|
| $Cu/SiO_2/Pt$ | 4 GΩ | 13 GΩ [a] | 5.8 GΩ | 3.3 [c] |
| $Ag/SiO_2/Pt$ | 0.4 GΩ | 0.9 GΩ [a] | 0.7 GΩ | 2.5 [c,d] |
| $Ag/GeS_{2.2}/Pt$ | 3 kΩ [b] | 3.4 kΩ | 27 kΩ | 1.13 [c,e] |
| $Ag/GeSe_{2.3}/Pt$ | 1 kΩ [b] | 1.3 kΩ | 4.5 kΩ | 1.3 [c,e] |
| Ag/AgI/Pt | 0.2 GΩ | 0.3 GΩ [a] | 0.6 GΩ | 1.3 [e] |
| $Cu/WO_x/Pt$ | 0.4 GΩ [b] | 1.1 GΩ | 0.6 GΩ | 2.75 [c] |
| $Pt/SrTiO_3/Ti$ | 1 MΩ | 1 GΩ [a] | ~ 1 MΩ | $\sim 10^3$ [f] |
| $Pt/Ta_2O_5/Ta$ | 10 kΩ | 10 MΩ [a] | ~ 10 kΩ | $\sim 10^3$ [f] |

[a] Short Circuit measurement (calculated by cell voltage and short circuit current)
[b] Impedance spectroscopy (using a PARSTAT 2273, input impedance $10^{13}$ Ω, ac-amplitude 20 mV, frequency range 2 MHz to 1 Hz)
[c] Hebb-Wagner measurement (refs [25, 26])
[d] emf slope (see Fig. 2c and eq. (11))
[e] emf method in $Ag/Electrolyte/C(I_2)$ cell
[f] Maximum transference number estimation based on the ratio of charge during switching and maximum amount of $O^{2-}$ ions in the cell which can be oxidized.

**Table S1.** Resistance values for the total ($R_{tot}$), the ion ($R_i$) the electronic ($R_e$) resistances, respectively and the ration $R_i/R_{tot}$. The methods used for determining the values are summarized in the trailer of the table. Please note that the electronic resistance $R_e$ for ECM cells in the OFF state is mainly determined by the leakage current $R_{leak}$.

The value of $R_i$ (and $R_{tot}$) and also the ratio $R_i/R_{tot}$ usually depends strongly on the chemical composition, on the amount of the dopant (e.g. Ag), electrolyte thickness, temperature, and gas atmosphere. Thus, the values we provide in the table are valid strictly for the corresponding system and may significantly vary if some of the parameters (e.g. amount of Ag ion doping) are changed.

### S12 Measurement Resolution and Accuracy

All measurements were performed in ultra low noise, dark and RF shielded measurement systems to prevent radio frequency interference (RFI) and electromagnetic interference (EMI) in accordance to ref. [27].

The triaxial measurement setup makes it possible to measure currents well below the pA range and significantly reduces the effective cable capacitance[28]. The measurement resolution has been checked to be within the device specification (Keithley 6430 Subfemto Remote SourceMeter, Keithley 617 Electrometer and Keithley 2636A SourceMeter). The current measurement resolution for short circuit measurements (compare Fig. 2d) is about 0.01 pA.

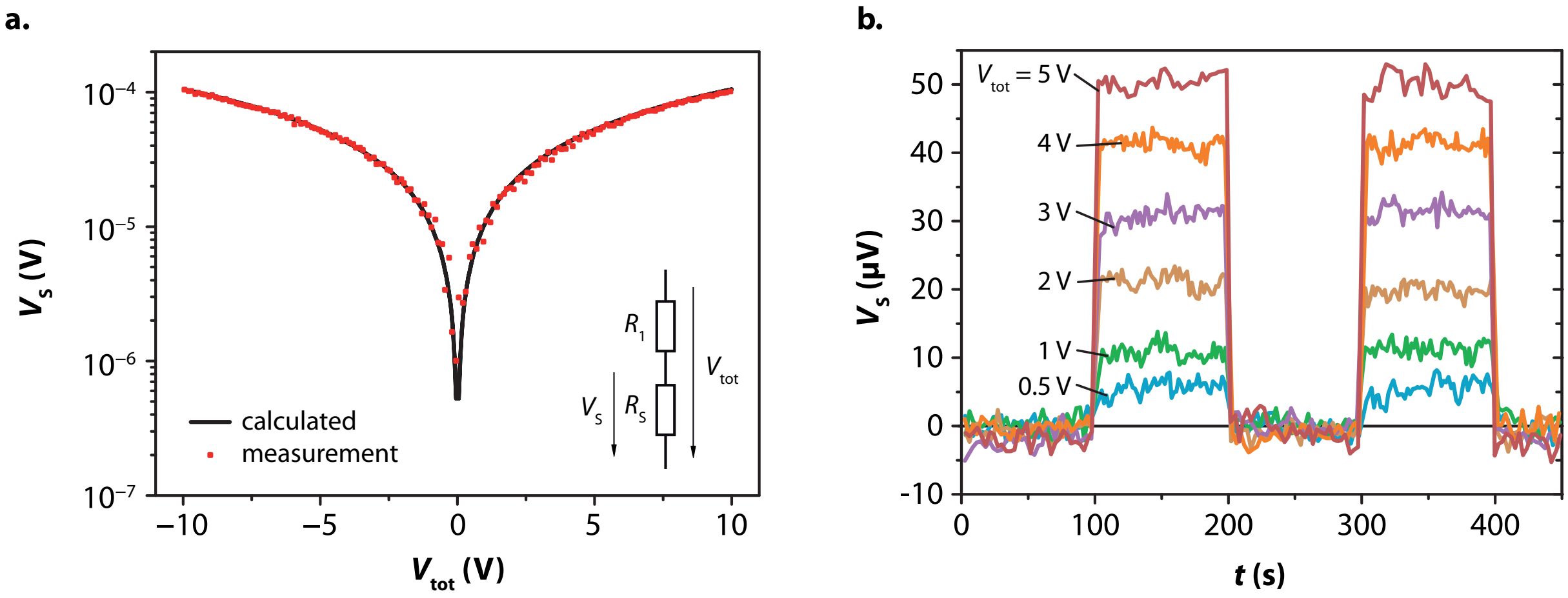

**Figure S12.1** Voltage measurement accuracy test measuring the voltage drop $V_S$ across a shunt resistor $R_S$ by variation of the applied voltage $V_{tot}$ ($R_S = 10\ \Omega$ and $R_1 = 10\ M\Omega$). **(a)** Small voltages even below 10 μV can be clearly measured and fit to the calculated voltage drop. **(b)** Voltage pulses with variation of the pulse amplitude $V_{tot}$ reveal a sufficient SNR (signal to noise ratio).

Three additional control experiments were conducted to exclude RFI or EMI effects as the origin of our emf voltage: (1) we transferred the measurement system into a RF shielded room and we repeated selected experiments. The results did not change within the measuring accuracy. (2) We tested a symmetric Pt/$GeSe_x$/Pt cell which should not show any significant cell voltage. In fact, the open cell voltage was in the noise level, i.e. no emf voltage has been detected. (3) Furthermore, we determined the voltage resolution limit of our system. Fig. S12.1a depicts the measured voltage drop across a shunt resistor $R_S$ compared to a calculated voltage drop. Fig. S12.1b shows the voltage over time for the given rectangular signal. The peak-to-peak noise level is approx. 7 μV, whereas the rms noise signal is 1 μV. As a consequence, voltages below 10 μV can be clearly measured. Since the voltage divider was placed in the same measurement system, the voltage stimulation steps and their response provide an additional proof that RFI or EMI effects can be excluded.

A thermoelectric effect as the origin of our emf voltage can be excluded because of the orders of magnitude different time constants. We measured the transient (steady-state open-cell) voltages from several minutes up to hours and even for days (e.g. for Cu/$SiO_2$/Pt, see Fig. S8.1b) while any temperature differences caused by prior switching vanish within nanoseconds to microseconds due to the small volume of the switching region. All emf measurements were performed at room temperature and after temperature gradients were equilibrated. Please note that emf voltages were measured in the OFF (or as deposited) state using high impedance source meters.

## S13 Experimental Summary

The following table contains a summary of most significant experiments within the context of this emf study.

| System | Measurement | Number of experiments |
|---|---|---|
| Ag/$SiO_2$ (30 nm … 50 nm)/Pt; d = 100 µm, 150 µm, 250 µm | Resistive Switching using *I*/*V* sweeps, SET/RESET experiments (examples shown or used in Fig. 2b, 2d, S3.1a, S5.1, S5.2 S9.1) | 320 |
| Ag/$SiO_2$ (50 nm)/Pt; d = 100 µm, 150 µm, 250 µm | Short circuit current ($V = 0$) after RESET (examples shown in Fig. 2d, S5.2) | 35 |
| Ag/$SiO_2$ (50 nm)/Pt; d = 100 µm, 150 µm, 250 µm | Cyclic Voltammetry (examples shown or used in Fig. 2b, 2c, S3.1b) | 36 |
| Ag/$SiO_2$ (50 nm)/Pt; d = 100 µm, 150 µm, 250 µm | Emf measurements (examples shown or used in Fig. 2b, 2c, 4a, S3.1) | 29 |
| Cu/$SiO_2$ (30 nm)/Pt; d = 100 µm | Resistive Switching using *I*/*V* sweeps, SET/RESET experiments (example shown in Fig. 4b) | 38 |
| Cu/$SiO_2$ (30 nm)/Pt; d = 100 µm | Emf measurements and Cyclic Voltammetry (examples shown or used in Fig. 4a, S3.1b, S8.1b) | 106 |
| Ag/AgI (30 nm … 50 nm)/Pt; $A_{Cell}$ = 3 µm x 3 µm (crossbar) | Emf measurements and Resistive Switching (example used in Fig. 4a) | 49 |
| TiN/$TiO_2$ (10 nm)/Ti; $A_{Cell}$ = 55 nm x 55 nm | Emf measurements and Resistive Switching (example used in Fig. S7.2) | 1 |
| Pt/$SrTiO_3$ (10 nm)/Ti; $A_{Cell} = 10^{-4}$ cm$^2$ | Resistive Switching (examples used or shown in Fig. 4a, S8.1a, S8.2a, S9.2) | 75 |
| Pt/$SrTiO_3$ (10 nm)/Ti; $A_{Cell} = 10^{-4}$ cm$^2$ | Emf measurements (example used in Fig. 4a, S8.1a, S8.2a) | 70 |
| Pt/$Ta_2O_5$ (10 nm)/Ti; $A_{Cell} = 10^{-4}$ cm$^2$ | Resistive Switching (examples used or shown in Fig. 4a, S8.2b) | 87 |
| Pt/$Ta_2O_5$ (10 nm)/Ti; $A_{Cell} = 10^{-4}$ cm$^2$ | Emf measurements (examples used or shown in Fig. 4a, S8.2b) | 80 |
| Cu/$WO_3$ (30 nm)/Pt; d = 50 µm | Emf measurements and Resistive Switching (used in Fig. 4a and Section S2) | 15 |
| Ag/GeS$_{2.2}$ (40 nm … 70 nm)/Pt; d = 100 µm | Emf measurements and Resistive Switching (used in Fig. 4a and S2.1) | 30 |
| Ag/GeSe$_{2.3}$ (40 nm … 70 nm)/Pt; d = 100 µm | Emf measurements and Resistive Switching (used in Fig. 4a, and section S2) | 24 |
| Ag/GeSe$_{2.3}$ (40 nm … 70 nm)/Pt; d = 100 µm | Optical analysis of dendrites (used in Fig. 3) | 4 |

**Table S2.** Summary of switching and emf experiments performed in the context of this study.

## References

1. Chua, L.O. Memristor-the missing circuit element. *IEEE Trans. Circuit* Theory CT-18, 507-519 (1971).

2. Chua, L.O. & Kang, S.M. Memristive devices and systems. *Proc.* IEEE 64, 209-223 (1976).

3. Strukov, D. B., Snider, G. S., Stewart, D. R. & Williams, R. S. The missing memristor found. Nature 453, 80-83 (2008).

4. Chua, L.O. Resistance switching memories are memristors. *Appl. Phys. A-Mater. Sci. Process.* 102, 765-783 (2011).

5. Cho, D.-Y., Valov, I., van den Hurk, J., Tappertzhofen, S. & Waser, R. Direct Observation of Charge Transfer in Solid Electrolyte for Electrochemical Metallization Memory. *Advanced* Materials 24, 4552-4556 (2012).

6. Tsuruoka, T., Terabe, K., Hasegawa, T. & Aono, M. Forming and switching mechanisms of a cation-migration-based oxide resistive memory. Nanotechnology 21, 425205/1-8 (2010).

7. Tappertzhofen, S., Menzel, S., Valov, I. & Waser, R. Redox Processes in Silicon Dioxide Thin Films using Copper Microelectrodes. *Appl. Phys. Lett.* 99, 203103/1-3 (2011).

8. Tappertzhofen, S., Mündelein, H., Valov, I. & Waser, R. Nanoionic transport and electrochemical reactions in resistively switching silicon dioxide. Nanoscale 4, 3040-3043 (2012).

9. Bard, A. & Faulkner, L. *Electrochemical Methods: Fundamentals and* Applications (John Wiley and Sons, New York, 2001).

10. Miao, F., Yang, J. J., Borghetti, J., Medeiros-Ribeiro, G. & Williams, R. S. Observation of two resistance switching modes in TiO2 memristive devices electroformed at low current. Nanotechnology 22, 254007/1-7 (2011).

11. Choi, S., Balatti, S., Nardi, F. & Ielmini, D. Size-dependent drift of resistance due to surface defect relaxation in conductive-bridge memory. *IEEE Electron Device Lett.* 33, 1189-91 (2012).

12. Valov, I., Waser, R., Jameson, J. R. & Kozicki, M. N. Electrochemical metallization memories-fundamentals, applications, prospects. Nanotechnology 22, 254003/1-22 (2011).

13. Waser, R., Menzel, S. & Bruchhaus, R., in: Nanoelectronics and Information Technology, 3rd edition (ed. R. Waser), *Wiley-VCH* (2012).

14. McBrayer, J. D., Swanson, R. M. & Sigmon, Y. W. Diffusion of metals in silicon dioxide. *J. Electrochem. Soc.* 133, 1242-6 (1986).

15. Noman, M., Jiang, W., Salvador, P. A., Skowronski, M. & Bain, J. A. Computational investigations into the operating window for memristive devices based on homogeneous ionic motion. *Appl. Phys. A - Mater. Sci. Process.* 102 (2011).

16. Kosacki, I., Rouleau, C. M., Becher, P. F., Bentley, J. & Lowndes, D. H. Nanoscale effects on the ionic conductivity in highly textured YSZ thin films. *Solid State* Ionics 176, 1319-1326 (2005).

17. Korte, C., Schichtel, N., Hesse, D. & Janek, J. Influence of interface structure on mass transport in phase boundaries between different ionic materials. *Monatshefte für Chemie - Chemical* Monthly 140, 1069-1080 (2009).

18. Maier, J. Nanoionics: ion transport and electrochemical storage in confined systems. *Nat. Mater.* 4, 805-815 (2005).

19. Maier, J. Thermodynamics of Nanosystems with a Special View to Charge Carriers. *Adv. Mater.* 21, 2571-2585 (2009).

20. Ilia Valov, & Kozicki, Michael N. Cation-based resistance change memory. *J. Phys. D Appl. Phys.* 46, 074005 (2013).

21. Simmons, J. G. Generalized Formula for the Electric Tunnel Effect between Similar Electrodes Separated by a Thin Insulating Film. *J. Appl. Phys.* 34, 1793-1803 (1963).

22. Menzel, S., Böttger, U. & Waser, R. Simulation of multilevel switching in electrochemical metallization memory cells. *J. Appl. Phys.* 111, 014501/1-5 (2012).

23. Tappertzhofen, S. *et al.* Capacity based Nondestructive Readout for Complementary Resistive Switches. Nanotechnology 22, 395203/1-7 (2011).

24. Chua, L., Sbitnev, V. & Kim, H. Hodgkin–Huxley Axon is Made of Memristors. *International Journal of Bifurcation and* Chaos 22, 1230011/1-48 (2012).

25. Carl Wagner, Proc. 7th Meet. Int. Comm. On Electrochem. Thermodynam. Kinet. *Butterworths,* London (1957).

26. Hebb, M. H. Electrical Conductivity of Silver Sulfide. *The Journal of Chemical* Physics 20, 185-190 (1952).

27. Keithley, *Low Level Measurements* Handbook (Keithley Instruments Inc., 2004).

28. Keithley, *Application Note 1671: Low Current* Measurements (Keithley Instruments Inc., 2012).

## Acknowledgement

The supply of Si wafers with nanostructured TiN bottom electrodes by IMEC, Leuven, for the result shown in Fig. S7.2 is gratefully acknowledged.